\documentclass[%
 reprint,
superscriptaddress,
onecolumn,
groupedaddress,
 amsmath,amssymb,
 aps,
 pra,
 nofootinbib,
]{revtex4-2}

\usepackage[caption=false]{subfig}

\usepackage{mathtools} 

\usepackage{graphicx}% Include figure files
\usepackage{dcolumn}% Align table columns on decimal point
\usepackage{bm}% bold math
\usepackage{xcolor}
\usepackage{soul}
\usepackage{mathtools}

\usepackage[hidelinks,colorlinks=true,linkcolor=blue,filecolor=blue,urlcolor=blue,citecolor=blue]{hyperref}

\usepackage{bbold}

\usepackage[normalem]{ulem}

\newcommand\blfootnote[1]{%
  \begingroup
  \renewcommand\thefootnote{}\footnote{#1}%
  \addtocounter{footnote}{-1}%
  \endgroup
}

\begin{document}

\preprint{APS/123-QED}

\title{Ratchet-mediated resetting: Current, efficiency, and exact solution}

\author{Connor Roberts}
\thanks{These authors contributed equally to this work}
\affiliation{Department of Mathematics, Imperial College London, London SW7 2AZ, United Kingdom}

\author{Emir Sezik}
\thanks{These authors contributed equally to this work}
\affiliation{Department of Mathematics, Imperial College London, London SW7 2AZ, United Kingdom}

\author{Eloise Lardet}
\affiliation{Department of Mathematics, Imperial College London, London SW7 2AZ, United Kingdom}

\date{\today}

%%%%%%%%%%%%%%%%%%%%%%%%%%%%%%%%%%%%%%%%%%%%%%%%%%

% ABSTRACT

%%%%%%%%%%%%%%%%%%%%%%%%%%%%%%%%%%%%%%%%%%%%%%%%%%

\begin{abstract}

We model an overdamped Brownian particle that is subject to resetting facilitated by a ratchet potential on a spatially periodic domain. This asymmetric potential switches on with a constant rate, but switches off again only upon the particle's first passage to a resetting point at the minimum of the potential.
Repeating this cycle sustains a non-equilibrium steady-state, as well as a directed steady-state current which can be harnessed to perform useful work. We derive exact analytic expressions for the probability densities of the free-diffusion and resetting phases, the associated currents for each phase, and an efficiency parameter that quantifies the return in current for given power input. These expressions allow us to fully characterise the system and obtain experimentally relevant results such as the optimal current and efficiency. Our results are corroborated by simulations, and have implications for experimentally viable finite-time resetting protocols. 

\end{abstract}

\maketitle

%%%%%%%%%%%%%%%%%%%%%%%%%%%%%%%%%%%%%%%%%%%%%%%%%%

% INTRODUCTION

%%%%%%%%%%%%%%%%%%%%%%%%%%%%%%%%%%%%%%%%%%%%%%%%%%

\section{Introduction}\label{Sec:Introduction}

\blfootnote{connor.roberts16@imperial.ac.uk}The field of stochastic resetting has received much attention since its inception over a decade ago \cite{evans2011, evans2020, gupta2022}.
In stochastic resetting, the recurrent restarting of a process from its initial state establishes internal probability currents and, remarkably, a non-equilibrium steady-state even if the underlying dynamics otherwise satisfy detailed balance. One of the most striking properties of stochastic resetting is that it can render the mean first-passage time of a diffusive search process finite \cite{evans2013, pal2016, pal2017}, thus offering an attractive way of improving transport properties \cite{evans2011a, alston2024}.
On top of its myriad variations and applications, such as to biology \cite{roldan2016, boyer2014}, computer  science \cite{ginsberg1993}, and finance \cite{cheng2000}, resetting has most recently been shown to improve the efficiency of Brownian heat engines \cite{lahiri2024}.

The prototypical model of stochastic resetting involves a Brownian particle subject to instantaneous resets to its initial position at a constant rate. This is in stark contrast to experimental implementations of stochastic resetting, where it takes a finite time to reset a particle back to its initial position \cite{tal-friedman2020, besga2020, faisant2021, goerlich2023}. With these recent experimental advances, others have sought to devise theoretical models that emulate finite-time resetting \cite{alston2022, pal2019, pal2020, bodrova2020resetting, ghosh2023, pal2019time, maso-puigdellosas2019, bodrova2020brownian, mercado-vasquez2020, mercado-vasquez2022, santra2021}.
Unlike instantaneous resetting, finite-time resetting can be mediated by a physical mechanism, such as an external potential \cite{gupta2020, gupta2021, biswas2023, olsen2024}. Aside from being more physically realistic, it was recently shown that potential-mediated resetting can actually expedite diffusive search processes compared to instantaneous resetting \cite{biswas2023, biswas2024}.

The probability currents generated through stochastic resetting naturally lead to questions about work extraction and the efficiency of resetting \cite{fuchs2016, gupta2022a, gupta2020work, busiello2020}. For instance, these models could be modified to generate useful work by introducing an underlying asymmetry into the resetting mechanism. Harnessing the asymmetry of an autonomous engine to generate work is an idea dating back to Feynman's ``ratchet and pawl'' machine \cite{feynman1963}. In a similar vein, models of molecular motors provide a mechanism for Brownian particles to exhibit a net drift \cite{julicher1997, reimann2002}, akin to the movement of motor proteins along filaments \cite{astumian2002, schliwa2003}. 
This directed motion can be brought about by the asymmetry of a ratchet potential, in combination with either potentials, external forces or states that fluctuate in time \cite{magnasco1993, astumian1994, doering1995}.
Alternatively, the particles themselves can self-propel, in which case their motion in an asymmetric potential readily breaks time-reversal symmetry, as has been the study of so-called ``active ratchets" in recent times \cite{angelani2011, zhen2022, su2023}.
Active ratchets are essentially autonomous engines that can be exploited to extract useful thermodynamic work \cite{pietzonka2019, roberts2023}.
Finding exact analytical solutions to these minimal models of non-equilibrium systems is an intense area of ongoing research \cite{malakar2018, razin2020, cocconi2020, garcia-millan2021, caraglio2022, roberts2022, alston2022, roberts2023}.

There has been a recent surge of numerical studies investigating resetting in combination with asymmetric potentials \cite{ghosh2023, fang2023, luo2022}. However, analytical studies are comparatively lacking, especially for experimentally relevant models. In this paper, we study the motion of an overdamped Brownian particle subject to finite-time resetting facilitated by a piecewise-linear ratchet potential on a periodic domain. The potential switches on with a constant rate, but switches off only upon the particle's first passage to a resetting point at the minimum of the potential, at which point the particle returns to a freely diffusive motion and restarts the cycle. This is in contrast to other models of fluctuating potential ratchets \cite{astumian1994, doering1995, reimann2002}, which have previously been used to model molecular motors \cite{julicher1997}. In these models, a time-varying ratchet potential switches between two or more states (such as ``on" and ``off"). In non-resetting ratchet models, this is controlled by some other form of systematic driving, e.g.\ switching between states with a constant rate. In the present model, the switching off being determined by the particle reaching the minimum of the potential means the particle can fully exploit the potential's height to further bias its motion through the ratchet's asymmetry, paving the way for an optimised current. This work brings together stochastic resetting and fluctuating potentials to provide a minimal finite-time resetting model that exhibits a net current. We derive the exact steady-state solution, backed up by simulation results, in an experimentally motivated stochastic resetting set-up. 

The remainder of the paper is organised as follows.
After introducing the model in Section~\ref{Sec:Model}, we derive the system's steady-state probability density in Section~\ref{Sec:ProbabilityDensities}. This is supplemented by a derivation of the time-dependent probability densities in Appendix~\ref{App:Sec:DensitySolutions}.
In Section~\ref{Sec:Current}, we turn our attention to the current, which, to the best of our knowledge, constitutes the first systematic analytic study of a steady-state net current generated through finite-time resetting.
The fact that we can generate a net current motivates us to study the energetic cost of this resetting in Section~\ref{sec:Efficiency}, and ultimately devise a measure of the efficiency of the current generation for given power input.
Throughout the paper, our analytical results are supported by extensive numerical simulations. Finally, we conclude and summarise our results in Section~\ref{Sec:Conclusion}.

%%%%%%%%%%%%%%%%%%%%%%%%%%%%%%%%%%%%%%%%%%%%%%%%%%

% MODEL

%%%%%%%%%%%%%%%%%%%%%%%%%%%%%%%%%%%%%%%%%%%%%%%%%%

\section{Model}\label{Sec:Model}

\begin{figure}
    \centering
    \includegraphics[width=0.75\linewidth]{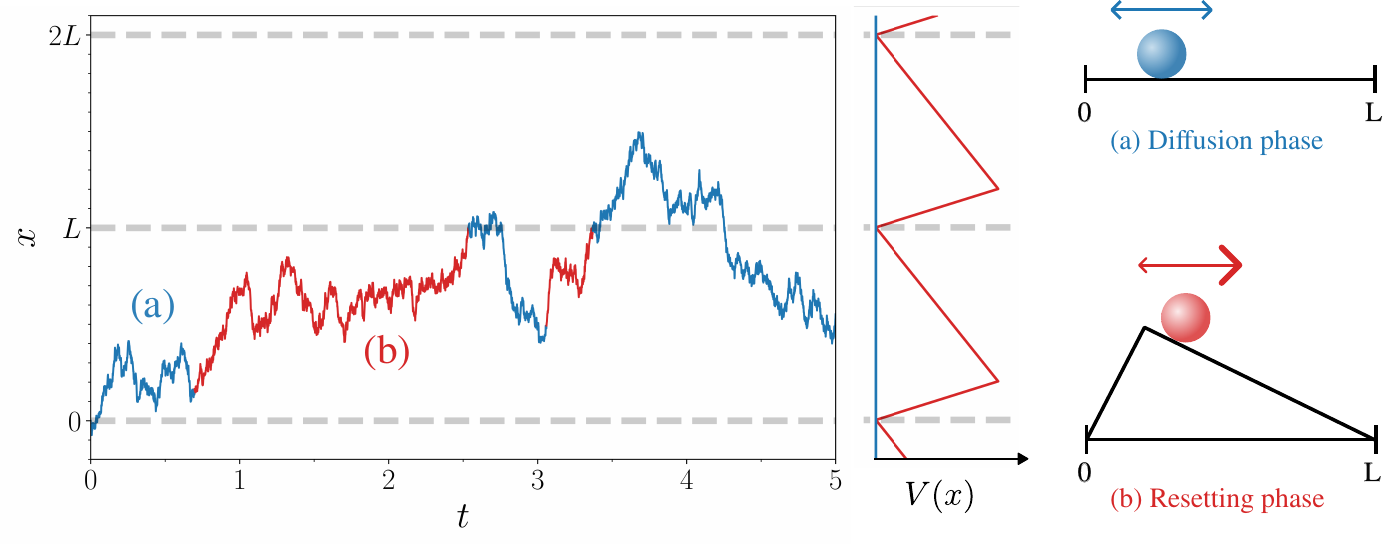}
    \caption{Example simulated trajectory from numerically integrating Eqs.~(\ref{eq:LangevinEquationDiffusion}) and (\ref{eq:LangevinEquationResetting}) for parameters $L=1$, $a=0.2$, $h=0.1$, $D_0 = 0.1$, and $r=2$, see Appendix~\ref{App:Sec:Simulations} for simulation details. Here, the trajectory is plotted on an extended domain for visualisation purposes; however, all results in the main text are calculated for a position $x$ wrapped back onto a periodic domain $x\in[0,L)$. The blue (respectively, red) sections of the line indicate when the particle is in the diffusion (respectively, resetting) phase. The ratchet potential $V(x)$, Eq.~(\ref{eq:ratchetpotential}), is visualised alongside the particle trajectory.
    The horizontal dashed lines indicate the resetting positions at $x=nL$ for $n\in \mathbb{Z}$. Resetting phases can start at any position $x$, but must end at these resetting points.
    }
    \label{fig:particle-trajectory}
\end{figure}

We consider an overdamped Brownian particle in one dimension which has two phases to its motion. In the ``diffusion phase", the particle freely diffuses and so its stochastic motion is governed by the following Langevin equation,
\begin{equation}
\label{eq:LangevinEquationDiffusion}
    \dot{x}_{D}(t) = \sqrt{2D_0} \xi(t),
\end{equation}
where $t$ is time, $D_0$ is the diffusion constant, $\xi(t)$ is a Gaussian white noise with zero mean $\xi(t) = 0$ and correlation $\langle \xi(t) \xi(t') \rangle = \delta(t-t')$, and the subscript $D$ indicates that motion is in the diffusion phase. The friction coefficient that would normally appear in Eq.~(\ref{eq:LangevinEquationDiffusion}) has been set to unity without loss of generality. 

While in the diffusion phase, a potential $V(x)$ with period $L$ is switched on with a Poissonian rate $r$, upon which the particle enters the ``resetting phase''. The potential is switched off again only upon the particle reaching a resetting point located at $x=nL$ for $n\in\mathbb{Z}$, which coincide with the potential minima. To obtain a model with non-vanishing steady-state density and current, we wrap the position $x$ onto a periodic domain, i.e.\ identify positions $x + nL$ with $x$, thus constraining the particle's motion to the interval $x\in[0,L)$. Due to these periodic boundary conditions, any reference henceforth to $x=0$ can be assumed to equally refer to $x=L$, and vice versa. 

By iterating the cycle of diffusion and resetting phases, a non-equilibrium steady-state is sustained from the constant injection of energy at the beginning of each resetting phase to increase the particle's potential energy. Denoting the period where the potential $V(x)$ is switched on with a subscript $R$, the Langevin equation for the particle while in the resetting phase is
\begin{equation}
\label{eq:LangevinEquationResetting}
    \dot{x}_{R}(t) = -V'\left(x_R(t)\right) + \sqrt{2D_0} \xi(t).
\end{equation}
We are interested in using the resetting potential $V(x)$ to rectify the particle's motion in order to generate a steady-state current, in which case the potential needs to impart a bias on the particle's motion. To render our system exactly solvable, we choose $V(x)$ to be a piecewise-linear asymmetric potential with an apex at $x=a$ \cite{roberts2023, astumian1994}, i.e.\ a ``ratchet potential'' defined by
\begin{equation}
\label{eq:ratchetpotential}
    V(x) = 
    \begin{cases}
      V^{[1]}(x) = \frac{h}{a}x, & 0 \leq x < a, \\
      V^{[2]}(x) = \frac{h}{L-a}(L-x), & a \leq x < L,
    \end{cases}
\end{equation}
where $h$ is the maximum height of the potential, occurring at $x=a$. Notably, this setup generalises instantaneous stochastic resetting, since the limit $h \to \infty$ corresponds to the particle being instantaneously reset to the origin $x=0$ when the potential is switched on. In the instantaneous resetting case, the parameter $a$ controls whether the particle instantaneously jumps a distance $x$ to the left, if $0 \leq x < a$, or a distance $L-x$ to the right, if $a \leq x < L$. The full model for finite $h$ is illustrated by an example stochastic trajectory in figure~\ref{fig:particle-trajectory}.

To demonstrate the rectification mechanism, consider the case where $a < L/2$. When the potential is switched on, the particle is more likely to be found in the larger $a \leq x < L$ region of the resetting potential, which imparts a positive drift velocity $-V^{[2]'}(x) = h/(L-a)$ on the particle  towards the origin $x=L^-$. Hence, on average, the particle experiences a positive drift while in the resetting phase. Since motion in the diffusion phase is symmetric, the particle's overall motion is rectified in the positive direction, corresponding to a positive net current. The equivalent negative current is obtained for a ratchet potential where the position of the apex is reflected across $x=L/2$, i.e.\ $a \to L-a$. 

Using a periodic asymmetric potential in combination with resetting to generate a net current was recently investigated in Ref.~\cite{ghosh2023}. However, in that work, the current was not induced by recurrent activation of the potential, as in the present study. Instead, the authors of Ref.~\cite{ghosh2023} obtained numerical results for a system in which a Brownian particle moves in a static asymmetric potential. The particle switches off its diffusion with a constant rate, leading to a deterministic return to the potential minimum proceeded by the particle restarting its diffusion. We argue that our present model is more experimentally accessible than this ``stop-and-start" model, as diffusion is not typically an internal degree of freedom that can be autonomously switched on or off, whereas an external potential can be readily modulated by an external controller. Furthermore, we obtain exact results for our model that can be used to analytically study experimentally relevant observables, such as the probability densities and currents for each phase, as well as the efficiency of the current generation. For completeness, we derive the analytic solution for the stop-and-start model in Appendix~\ref{App:Sec:DeterministicRatchetModel}.

%%%%%%%%%%%%%%%%%%%%%%%%%%%%%%%%%%%%%%%%%%%%%%%%%%

% Probability densities

%%%%%%%%%%%%%%%%%%%%%%%%%%%%%%%%%%%%%%%%%%%%%%%%%%

\section{Probability densities}\label{Sec:ProbabilityDensities}

The dynamics of the system can be equivalently encoded in the probability densities $P_{D/R}(x,t)$ to observe the particle in the diffusion/resetting phase at a position $x$ at time $t$. Since the ratchet potential, Eq.~(\ref{eq:ratchetpotential}), is piecewise linear, it is convenient to separate the probability density for the resetting phase into separate components for each region. Here and throughout, we distinguish these separate components with a superscript $[i]$, where $i \in \{1,2\}$, i.e.\
\begin{equation}\label{eq:Densities_Separate}
    P_R(x,t) = 
    \begin{cases}
      P_R^{[1]}(x,t), & 0 \leq x < a \\
      P_R^{[2]}(x,t), & a \leq x < L
    \end{cases}.
\end{equation}
The probability densities obey the following coupled Fokker-Planck equations,
\begin{subequations}
    \label{eq:FPequation}
    \begin{align}
         \frac{\partial P_D(x,t)}{\partial t} &= -\frac{\partial J_D(x,t)}{\partial x}- rP_D(x,t) + \left(J_R^{[2]}(L^{-},t) - J_R^{[1]}(0^{+},t)\right)\delta(x) ,\label{eq:FP_Diffusion} \\
         \frac{\partial P_R^{[1]}(x,t)}{\partial t} &= -\frac{\partial J_R^{[1]}(x,t)}{\partial x} + r P_D(x,t) +  J_R^{[1]}(0^{+},t)\delta(x) , \label{eq:FP_Resetting1} \\
         \frac{\partial P_R^{[2]}(x,t)}{\partial t} &= -\frac{\partial J_R^{[2]}(x,t)}{\partial x} + r P_D(x,t) -  J_R^{[2]}(L^{-},t)\delta(x) , \label{eq:FP_Resetting2}
    \end{align}
\end{subequations}
where $L^-$ (respectively, $0^+)$ is the limit of $x$ approaching $L$ (respectively, $0$) from below (respectively, above), and the currents for each phase are given by
\begin{subequations}
    \label{eq:currents}
    \begin{align}
         J_D (x,t) &= -D_0\frac{\partial P_D(x,t)}{\partial x} ,\label{subeq:Current_DiffusionPhase} \\
         J_R^{[i]} (x,t) &= -D_0\frac{\partial P_R^{[i]}(x,t)}{\partial x} -  V'^{[i]} P_R^{[i]}(x,t) , \label{subeq:Current_ResettingPhase}
    \end{align}
\end{subequations}
where each $V'^{[i]}$ is spatially independent, and $J_R^{[2]}(L^{-},t) - J_R^{[1]}(0^{+},t)$ is the net flux directed into the origin $x = 0$ in the resetting phase, which quantifies the rate at which the particle arrives at the resetting point. We always have $J_R^{[2]}(L^-,t) \geq 0$ and $J_R^{[1]}(0^+,t) \leq 0$, regardless of the system parameters, since the individual terms of Eq.~(\ref{subeq:Current_ResettingPhase}) are both positive for $J_R^{[2]}(L^-,t)$, and both negative for $J_R^{[1]}(0^+,t)$, as will become clear once $P_R(x,t)$ is derived later. Hence, the term $(J_R^{[2]}(L^{-},t) - J_R^{[1]}(0^{+},t)) \delta(x)$ in Eq.~(\ref{eq:FPequation}) always represents a source term for the diffusion phase, while the terms $-J_R^{[2]}(L^{-},t)\delta(x)$ and $J_R^{[1]}(0^{+},t)\delta(x)$ always represent sink terms for each region in the resetting phase.

Upon summing the probabilities in Eq.~(\ref{eq:FPequation}), we recover the continuity equation 
\begin{equation}\label{eq:continuityeq}
    \frac{\partial P(x,t)}{\partial t} = -\frac{\partial J(x,t)}{\partial x},
\end{equation}
where $P(x,t) = P_D(x,t) + P_R(x,t)$ is the overall probability density, $J(x,t) = J_D(x,t) + J_R(x,t)$ is the net current, and recall that $P_R(x,t)$ is composed of components $P_{R}^{[i]}(x,t)$ that are valid only for specific regions of space as per Eq.~(\ref{eq:Densities_Separate}). Hence, the sink term $-rP_D(x,t)$ from Eq.~(\ref{eq:FP_Diffusion}) is cancelled in Eq.~(\ref{eq:continuityeq}) by the source terms $rP_D(x,t)$ from Eqs.~(\ref{eq:FP_Resetting1}) and (\ref{eq:FP_Resetting2}) in a piecewise manner, first by that of Eq.~(\ref{eq:FP_Resetting1}) in the region $x\in [0,a)$, and then by that of Eq.~(\ref{eq:FP_Resetting2}) in the region $x\in [a,L)$.

Our setup generalises the mod-linear resetting model of Ref.~\cite{gupta2021} by allowing the resetting potential to be asymmetric and periodic in order to generate a non-zero net current $J(x,t)$. Thus, our model reduces to that of Ref.~\cite{gupta2021} by setting $a = L/2$ while taking $L \to \infty$ and $h \to \infty$ in such a way that the gradients $V'^{[i]}$ remain constant. The periodicity of the potential imposes the following periodic boundary conditions,
\begin{subequations}\label{eq:BC_origin}
    \begin{align}
    P_D(0,t) &= P_D(L,t), \label{subeq:PBC_Diffusion}\\
    P_R^{[1]}(0,t) &= P_R^{[2]}(L,t) = 0, \label{subeq:PBC_Resetting}
    \end{align}
\end{subequations}
where Eq.~(\ref{subeq:PBC_Resetting}) also imposes that a particle in the resetting phase can never be found at the origin because it would instantly transmute to the diffusion phase.

We are interested in the steady-state behaviour of the system, for which the left-hand sides of Eq.~(\ref{eq:FPequation}) vanish, i.e.\ $\lim_{t\to\infty} \partial_t P_{D,R}(x,t) = 0$. Henceforth, we will represent all steady-state quantities by dropping the `$t$' from their arguments, e.g.\ $P_D(x) \equiv \lim_{t\to\infty}P_D(x,t)$. The steady-state net current $J$ is independent of position as a result of the steady-state condition $\partial_t P(x) = -\partial_x J = 0$, where $P(x) = P_D(x) + P_R(x)$ is the overall probability density at stationarity.

We now recast the dynamics in dimensionless form. This will allow us to explore the full parameter space more efficiently by reducing the description of the system from five dimensionful parameters $h,a,L,r,D_0$, to three dimensionless parameters, to be specified below. Defining the typical diffusion length scale between resets $\sigma \coloneq \sqrt{D_0 / r}$ and the average time for the potential to switch on $\tau \coloneq 1/r$ as our units of distance and time, respectively, the dimensionless Fokker-Planck equation at stationarity is
\begin{subequations}
    \label{eq:FP_dimensionless}
    \begin{align}
         0 &= \frac{\partial^2 \tilde{P}_D(\tilde{x})}{\partial \tilde{x}^2}- \tilde{P}_D(\tilde{x}) + \left(\tilde{J}_R^{[2]}(\ell^{-}) - \tilde{J}_R^{[1]}(0^{+})\right) \delta(\tilde{x}) ,\label{subeq:FP_diffusion_dimensionless} \\
         0 &= \frac{\partial^2 \tilde{P}_R^{[1]}(\tilde{x})}{\partial \tilde{x}^2}  + \tilde{V}'^{[1]}\frac{\partial \tilde{P}_R^{[1]}(\tilde{x})}{\partial \tilde{x}} + \tilde{P}_D(\tilde{x}) +\tilde{J}_R^{[1]}(0^{+}) \delta(\tilde{x}),
         \label{subeq:FP_resetting_dimensionless1} \\
         0 &= \frac{\partial^2 \tilde{P}_R^{[2]}(\tilde{x})}{\partial \tilde{x}^2}  + \tilde{V}'^{[2]}\frac{\partial \tilde{P}_R^{[2]}(\tilde{x})}{\partial \tilde{x}} + \tilde{P}_D(\tilde{x}) - \tilde{J}_R^{[2]}(\ell^{-}) \delta(\tilde{x}),
         \label{subeq:FP_resetting_dimensionless2}
    \end{align}
\end{subequations}
where we have represented non-dimensionalised quantities with a tilde, e.g.\ $\tilde{x} = x/\sigma$, $\tilde{P}_{D,R} = \sigma P_{D,R}$, and $\tilde{J}_R = \tau J_R$. The gradients of the potential in dimensionless form are given by
\begin{equation}
\label{eq:RatchetGradient_dimensionless}
    \tilde{V}'(\tilde{x}) = 
    \begin{cases}
      \tilde{V}'^{[1]} = \frac{\kappa}{\delta \ell}, & 0 \leq \tilde{x} <  \delta \ell \\
      \tilde{V}'^{[2]} = -\frac{\kappa}{(1-\delta)\ell}, &  \delta \ell \leq \tilde{x} < \ell
    \end{cases},
\end{equation}
where $\ell \coloneq L/\sigma$ is the rescaled system length, and we have introduced relevant dimensionless parameters, namely the coupling strength $\kappa \coloneq h/D_0$ of the particle to the potential, and the potential asymmetry parameter $\delta \coloneq a/L \in [0,1]$, i.e.\ the position of the ratchet apex for an interval normalised to unit length. In dimensionless form, the periodic boundary conditions (\ref{eq:BC_origin}) become
\begin{subequations}\label{eq:BC_dimensionless}
    \begin{align}
    \tilde{P}_D(0,t) &= \tilde{P}_D(\ell,t), \label{subeq:PBC_Diffusion_dimensionless}\\
    \tilde{P}_R^{[1]}(0,t) &= \tilde{P}_R^{[2]}(\ell,t) = 0. \label{subeq:PBC_Resetting_dimensionless}
    \end{align}
\end{subequations}

\begin{figure}
    \centering
    \includegraphics[width=1.0\linewidth]{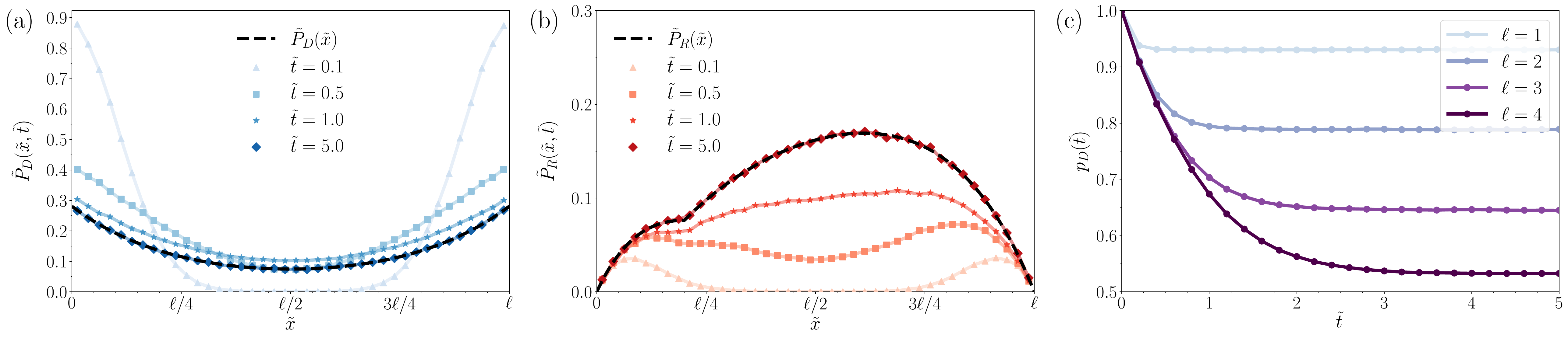}
    \caption{(a)-(b) Time evolution of the probability densities for each phase $\tilde{P}_{D,R}(\tilde{x},\tilde{t})$, as determined from simulations. Simulation parameters are $\ell=4$, $\kappa=1$, $\delta=0.2$ for a particle initialised at the origin $\tilde{x} = 0$ at time $\tilde{t}=0$, i.e.\ $\tilde{P}_D(\tilde{x},0) = \delta(\tilde{x})$. Full simulation details are given in Appendix~\ref{App:Sec:Simulations}. The time-dependent densities converge to the theoretical steady-state densities, $\tilde{P}_D(\tilde{x})$ and $\tilde{P}_R(\tilde{x})$, given in Eqs.~(\ref{eq:ProbabilityDensity_Diffusion_Solution}) and (\ref{eq:ProbabilityDensity_Resetting_Solution}), respectively. (c) Probability of being in the diffusion phase $p_D(\tilde{t})$ as a function of time for various values of $\ell$, as obtained from simulations for a single particle initialised at the origin $\tilde{x}=0$ in the diffusion phase at time $\tilde{t}=0$, with $\kappa=1$ and $\delta=0.2$.}
    \label{fig:TimeDependentProbabilityDensities}
\end{figure}

To obtain the steady-state probability density for the diffusion phase $\tilde{P}_D(\tilde{x})$, we first solve Eq.~(\ref{subeq:FP_diffusion_dimensionless}) away from the resetting point, i.e.\ in the region $\tilde{x} \in (0,\ell)$, where the term proportional to the Dirac delta function vanishes. This results in
\begin{equation}\label{eq:ProbabilityDensity_Diffusion_UnspecifiedConstants}
    \tilde{P}_D(\tilde{x}) = A_D e^{\tilde{x}} + B_D e^{- \tilde{x}}.
\end{equation}
By integrating Eq.~(\ref{subeq:FP_diffusion_dimensionless}) over the region $\tilde{x} \in [-\epsilon, \epsilon]$, recalling that $-\epsilon$ is identified with $\ell - \epsilon$, and then taking $\epsilon \to 0^+$, we obtain the following jump condition at the origin
\begin{equation}
    \label{eq:jumpcondition_D_main}
    \left. \frac{\partial \tilde{P}_D(\tilde{x})}{\partial \tilde{x}}\right|_{\tilde{x} \to 0^{+}} - \left. \frac{\partial \tilde{P}_D(\tilde{x})}{\partial \tilde{x}}\right|_{\tilde{x} \to \ell^{-}} = -\left(\tilde{J}_R^{[2]}(\ell^{-}) - \tilde{J}_R^{[1]}(0^{+})\right).
\end{equation}
Using this jump condition (\ref{eq:jumpcondition_D_main}) in combination with the periodic boundary condition (\ref{subeq:PBC_Diffusion_dimensionless}), we can fix the constants $A_D$ and $B_D$ in Eq.~(\ref{eq:ProbabilityDensity_Diffusion_UnspecifiedConstants}), eventually obtaining
\begin{equation}\label{eq:ProbabilityDensity_Diffusion_Solution}
    \tilde{P}_D(\tilde{x}) = \frac{p_D}{2}\frac{\cosh(\tilde{x}-\frac{\ell}{2})}{\sinh(\frac{\ell}{2})},
\end{equation}
where we have integrated Eq.~(\ref{eq:ProbabilityDensity_Diffusion_Solution}) over the entire interval $\tilde{x} \in [0, \ell]$ to yield $\tilde{J}_R^{[2]}(\ell^{-}) - \tilde{J}_R^{[1]}(0^{+}) = p_D$, where
\begin{equation}
    \label{eq:ProbabilityDiffusionPhase}
    p_{D} = \int_0^\ell d\tilde{x}~\tilde{P}_{D}(\tilde{x})
\end{equation}
is the probability to be in the diffusion phase at stationarity. We similarly define $p_R = p_{R}^{[1]} + p_{R}^{[2]}$ as the probability to be in the resetting phase at stationarity, where
\begin{subequations}\label{eq:ProbabilityResettingPhase}
    \begin{align}
    p_{R}^{[1]} &= \int_{0}^{\delta\ell} d\tilde{x}~\tilde{P}_{R}^{[1]}(\tilde{x}), \label{eq:ProbabilityResettingPhase_Region1}\\
    p_{R}^{[2]} &= \int_{\delta\ell}^{\ell} d\tilde{x}~\tilde{P}_{R}^{[2]}(\tilde{x}), \label{eq:ProbabilityResettingPhase_Region2}
    \end{align}
\end{subequations}
are the probabilities for the particle to be found in the different regions of the ratchet and in the resetting phase.

\begin{figure}
    \centering
    \includegraphics[width=0.7\linewidth]{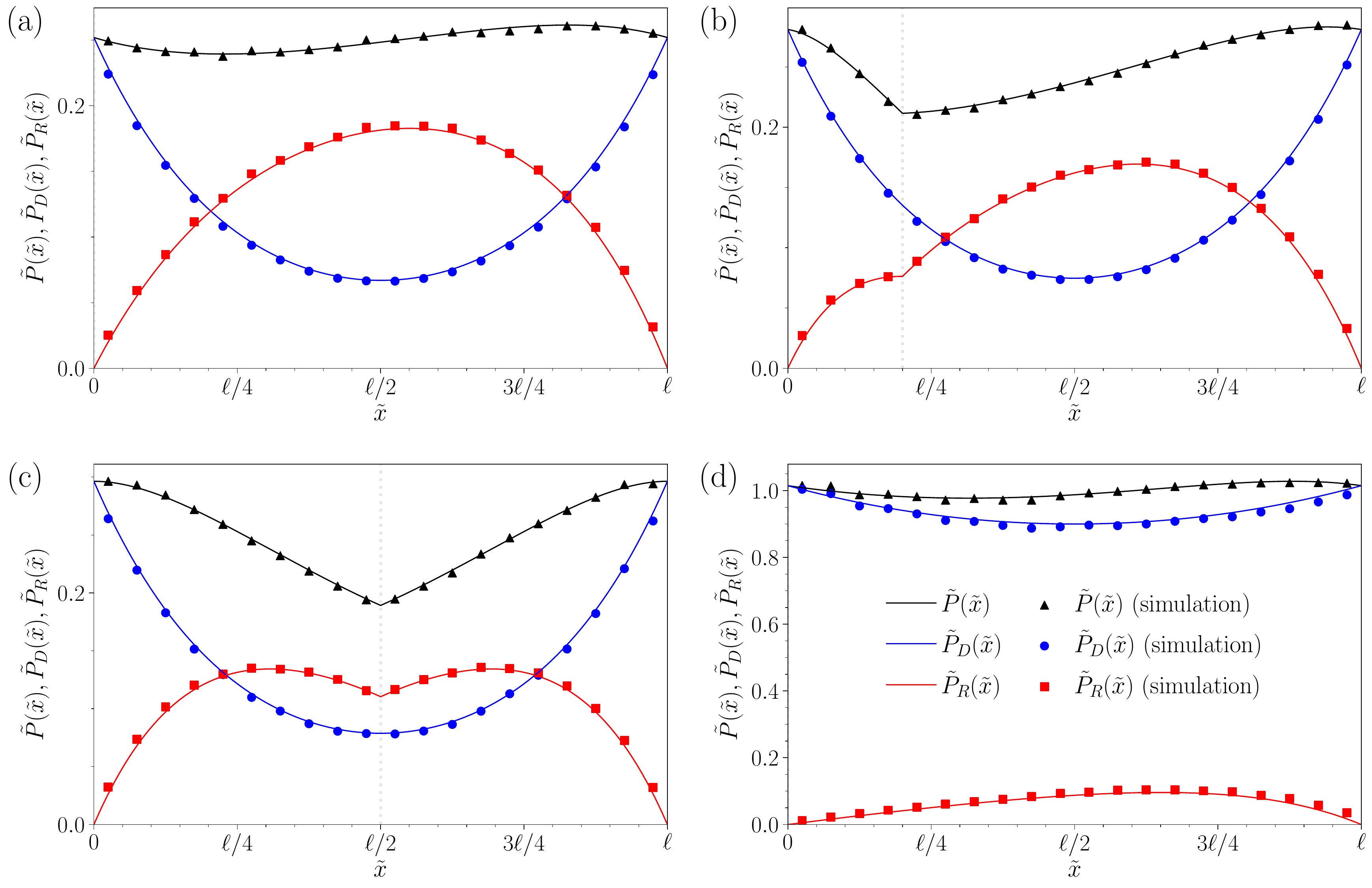}
    \caption{Steady-state probability densities $\tilde{P}_{D}(\tilde{x})$, Eq.~(\ref{eq:ProbabilityDensity_Diffusion_Solution}), $\tilde{P}_{R}(\tilde{x})$, Eq.~(\ref{eq:ProbabilityDensity_Resetting_Solution}), and $\tilde{P}(\tilde{x}) = \tilde{P}_{D}(\tilde{x}) + \tilde{P}_{R}(\tilde{x})$, as a function of particle position $\tilde{x}$ for parameters (a) $\delta = 0$, $\ell = 4$, $\kappa = 1$, (b) $\delta = 0.2$, $\ell = 4$, $\kappa = 1$, (c) $\delta = 0.5$, $\ell = 4$, $\kappa = 1$, and (d) $\delta = 0$, $\ell = 1$, $\kappa = 4$. Vertical dotted lines indicate the position of the ratchet apex $\tilde{x} = \delta\ell$ (equivalent to $x=a$) for the cases where $\delta > 0$. Simulation data are plotted as symbols and are in excellent agreement with the theoretical results. Full simulation details are given in Appendix~\ref{App:Sec:Simulations}.}
    \label{fig:ProbabilityDensitiesVsPosition}
\end{figure}

We now proceed to obtain the steady-state probability density for the resetting phase $\tilde{P}_R(\tilde{x})$. Due to the discontinuity in the potential at the apex $\tilde{x} = \delta \ell$, we also have continuity of the probability density and current for the resetting phase at this position, i.e.\
\begin{subequations}\label{eq:Continuity_dimensionless_Resetting}
    \begin{align}
    \tilde{P}_R^{[1]}(\delta\ell) &= \tilde{P}_R^{[2]}(\delta \ell), \label{subeq:Continuity_Resetting_dimensionless}\\
    \tilde{J}_R^{[1]}(\delta\ell) &= \tilde{J}_R^{[2]}(\delta\ell), \label{subeq:CurrentContinuity_Resetting_dimensionless}
    \end{align}
\end{subequations}
where the current for the resetting phase is defined in Eq.~(\ref{subeq:Current_ResettingPhase}). We also have the jump condition at the origin,
\begin{equation}
    \label{eq:jumpcondition_R_main}
    \left. \frac{\partial \tilde{P}_R^{[1]}(\tilde{x})}{\partial \tilde{x}}\right|_{\tilde{x} \to 0^{+}} - \left. \frac{\partial \tilde{P}_R^{[2]}(\tilde{x})}{\partial \tilde{x}}\right|_{\tilde{x} \to \ell^{-}} = \tilde{J}_R^{[2]}(\ell^{-}) - \tilde{J}_R^{[1]}(0^{+}) = p_D.
\end{equation}
As above, we first solve Eqs.~(\ref{subeq:FP_resetting_dimensionless1})-(\ref{subeq:FP_resetting_dimensionless2}) in the region $\tilde{x} \in (0,\ell)$, where the terms proportional to the Dirac delta function vanish. After inserting the solution for $\tilde{P}_D(\tilde{x})$, Eq.~(\ref{eq:ProbabilityDensity_Diffusion_Solution}), into Eqs.~(\ref{subeq:FP_resetting_dimensionless1})-(\ref{subeq:FP_resetting_dimensionless2}) and then solving, we eventually obtain
\begin{equation}\label{eq:ProbabilityDensity_Resetting_Solution}
    \tilde{P}_R(\tilde{x}) = 
    \begin{cases}
      \frac{p_D}{2} \frac{ \delta \ell}{e^{\ell} - 1}\left( \frac{e^{\ell - \tilde{x}}}{\lambda^{[1]}_{-}} - \frac{e^{\tilde{x}}}{\lambda^{[1]}_{+}}\right) + A_R^{[1]} e^{-\frac{\kappa \tilde{x}}{ \delta \ell}} + B_R^{[1]}, & 0 \leq \tilde{x} < \delta\ell \\
      \frac{p_D}{2} \frac{ (1 - \delta) \ell}{e^{\ell} - 1}\left( \frac{e^{\tilde{x}}}{\lambda^{[2]}_{-}} - \frac{e^{\ell - \tilde{x}}}{\lambda^{[2]}_{+}}\right) + A_R^{[2]} e^{\frac{\kappa \tilde{x} }{ (1 - \delta) \ell}} + B_R^{[2]},  & \delta\ell \leq \tilde{x} < \ell
    \end{cases},
\end{equation}
where we have written the solution in terms of the compact notation $\lambda^{[1]}_{\pm} = \kappa \pm  \delta\ell$, $\lambda^{[2]}_{\pm} = \kappa \pm  (1 - \delta)\ell$. Equation~(\ref{eq:ProbabilityDensity_Resetting_Solution}) is valid only when both $\lambda_{-}^{[1]} \neq 0$ and $\lambda_{-}^{[2]} \neq 0$. We provide the solutions for the edge cases (i) $\lambda_{-}^{[1]} = 0 \neq \lambda_{-}^{[2]}$, (ii) $\lambda_{-}^{[2]} = 0 \neq \lambda_{-}^{[1]}$, and (iii) $\lambda_{-}^{[1]} = \lambda_{-}^{[2]} = 0$ at the end of Appendix~\ref{App:Sec:DensitySolutions}. The constants $A_R^{[1]},A_R^{[2]},B_R^{[1]},B_R^{[2]}$ in Eq.~(\ref{eq:ProbabilityDensity_Resetting_Solution}) are fixed by the periodic boundary condition (\ref{subeq:PBC_Resetting_dimensionless}), continuity conditions (\ref{eq:Continuity_dimensionless_Resetting}), and the jump condition (\ref{eq:jumpcondition_R_main}). The diffusion-phase probability $p_D$ can be fixed in terms of the system parameters $\ell,\kappa,\delta$ through the normalisation condition,
\begin{equation}
    \label{eq:NormalisationCondition}
    p_D + p_R^{[1]} + p_R^{[2]} = 1,
\end{equation}
where the probabilities for the particle to be found in each phase $p_{D,R}$ are defined in Eqs.~(\ref{eq:ProbabilityDiffusionPhase})-(\ref{eq:ProbabilityResettingPhase}). In principle, applying all of the aforementioned conditions yields closed-form solutions for the densities $\tilde{P}_D(\tilde{x})$, Eq.~(\ref{eq:ProbabilityDensity_Diffusion_Solution}), and $\tilde{P}_R(\tilde{x})$, Eq.~(\ref{eq:ProbabilityDensity_Resetting_Solution}). However, we find the full expressions are too cumbersome to write down here, in particular that of the overall normalisation $p_D$. Nevertheless, it turns out that in certain (relevant) limits, other observables such as the net current $\tilde{J}$ are compact enough to display explicitly in terms of the system parameters. We demonstrate this in Sec.~\ref{Sec:Current}.

As noted in Refs.~\cite{gupta2020, gupta2021}, an alternative way to fix $p_D$ is to relate the probability of being in either phase to the mean first-passage time $T_R(x_0)$ for the particle to reach the origin $x=0$ (or $x=L$) from a position $x_0$ in the resetting phase. Specifically,
\begin{equation}
    \label{eq:NormalisationCondition_MFPT}
    p_D = \frac{\tau}{\tau + \int_0^L dx_0 ~T_R(x_0)P(x_0|D)},
\end{equation}
where $\tau = 1/r$ is the average time spent in the diffusion phase in a single cycle, and $P(x_0|D) = P_D(x_0)/p_D$ (the probability to be at position $x_0$ \textit{given} the particle is in the diffusion phase) provides the weighting for initially being found at position $x_0$ in the resetting phase, such that $ \int_0^L dx_0 ~T_R(x_0)P(x_0|D)$ is the average time spent in the resetting phase each cycle. However, Eqs.~(\ref{eq:NormalisationCondition}) and (\ref{eq:NormalisationCondition_MFPT}) are not independent conditions, since we show in Appendix~\ref{app:sec:RelatingNormalisationConditions} that Eq.~(\ref{eq:NormalisationCondition_MFPT}) can be derived from Eq.~(\ref{eq:NormalisationCondition}) using only the definition of the mean first-passage time. For completeness, we derive the mean first-passage time $T_R(x_0)$ in Appendix~\ref{App:Sec:ResettingMFPT}. The time-dependent behaviour of $p_D(t)$ is plotted in figure~\ref{fig:TimeDependentProbabilityDensities}c, where it can be seen that $p_D(t)$ saturates to a steady-state value $p_D$ determined by Eq.~(\ref{eq:NormalisationCondition}) or, equivalently, Eq.~(\ref{eq:NormalisationCondition_MFPT}).

In addition to the steady-state solutions derived in this section, we demonstrate how one can derive the fully time-dependent Laplace-space solutions in Appendix \ref{App:Sec:DensitySolutions}. The time-dependent densities, $\tilde{P}_D(\tilde{x},\tilde{t})$ and $\tilde{P}_R(\tilde{x},\tilde{t})$, are plotted in figures~\ref{fig:TimeDependentProbabilityDensities}a and \ref{fig:TimeDependentProbabilityDensities}b, respectively. It can be seen that the time-dependent densities converge to the expressions for the theoretical steady-state densities, $\tilde{P}_D(\tilde{x})$ and $\tilde{P}_R(\tilde{x})$, given in Eqs.~(\ref{eq:ProbabilityDensity_Diffusion_Solution}) and (\ref{eq:ProbabilityDensity_Resetting_Solution}), respectively.

The solutions for the steady-state densities $\tilde{P}_D(\tilde{x})$, Eq.~(\ref{eq:ProbabilityDensity_Diffusion_Solution}), and $\tilde{P}_R(\tilde{x})$, Eq.~(\ref{eq:ProbabilityDensity_Resetting_Solution}), are plotted in figure~\ref{fig:ProbabilityDensitiesVsPosition} alongside simulation results for a few different sets of parameters. Data from the simulations are in excellent agreement with the theoretical results. We observe that the probability density in the diffusion phase $\tilde{P}_D(\tilde{x})$ has a symmetric exponential form about the origin $x=0$, as is typical for a resetting model.  The diffusion-phase density is also periodic, as required by the boundary conditions, with a minimum at the centre of the ratchet potential $\tilde{x}=\ell/2$, such that it is also symmetric about $\tilde{x}=\ell/2$. The shape of the probability density in the resetting phase $\tilde{P}_R(\tilde{x})$ heavily depends on the position of the ratchet potential's apex $\delta\ell$. In particular, the resetting density is symmetric if the potential is symmetric, i.e.\ $\delta = 1/2$, see figure~\ref{fig:ProbabilityDensitiesVsPosition}c, but is otherwise asymmetric and generally has a prominent corner at the apex as a result of the different potential gradients either side of $\tilde{x} = \delta \ell$.

\begin{figure}
    \centering
    \includegraphics[width=\linewidth, trim = 0.5cm 0.4cm 0.2cm 0.5cm, clip]{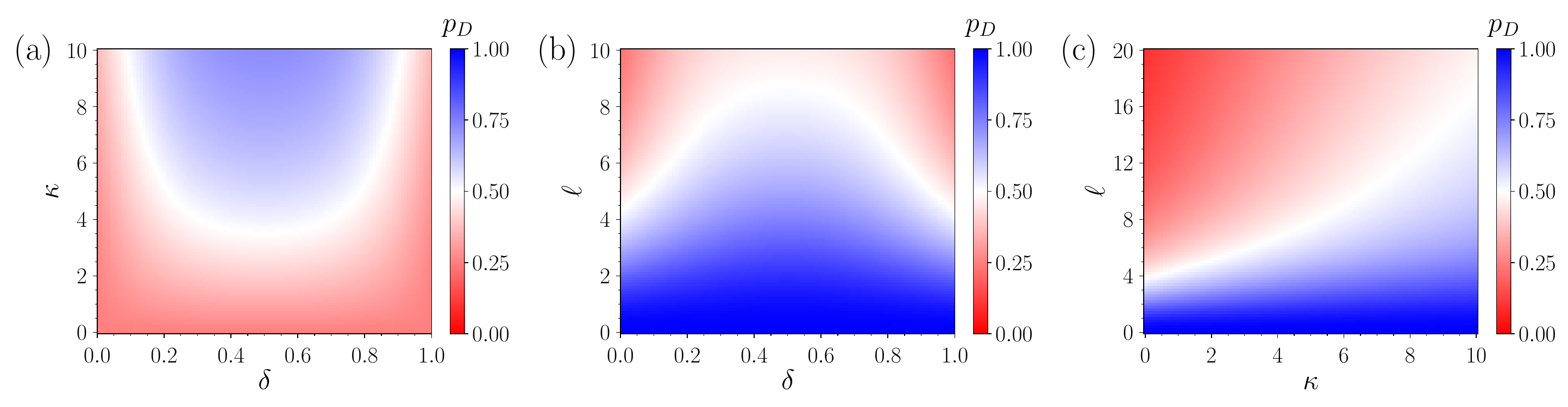}
    \caption{Steady-state probability to be in the diffusion phase $p_D$, Eq.~(\ref{eq:ProbabilityDiffusionPhase}), as a function of the dimensionless parameters $\ell$, $\kappa$, and $\delta$. The constant parameter in each subfigure is (a) $\ell = 8$, (b) $\kappa = 4$, and (c) $\delta = 0.2$. 
    }
    \label{fig:DiffusionProbabilityVsParameters}
\end{figure}

In figure~\ref{fig:DiffusionProbabilityVsParameters}, we plot the probability of the particle being in the diffusion phase at stationarity, Eq.~(\ref{eq:ProbabilityDiffusionPhase}). This captures the competition between the relevant timescales of the system, namely the average times spent in the diffusion and resetting phases, as illustrated by Eq.~(\ref{eq:NormalisationCondition_MFPT}). The particle has a greater chance of being in the diffusion phase for larger $\kappa$ and smaller $\ell$. Larger $\kappa$ indicates a stronger coupling to the potential, and therefore the potential resets the particle to the diffusion phase quicker, thus spending less time in the resetting phase and more time in the diffusion phase. In fact, $p_D \to 1$ as $\kappa \to \infty$. Similarly, smaller $\ell$ indicates the particle has been able to explore more of the interval between resetting phases, i.e.\ the density is more equilibrated between resets, and so the particle spends a greater proportion of its time in the diffusion phase. As above, $p_D \to 1$ as $\ell \to 0$.

In figure~\ref{fig:DiffusionProbabilityVsParameters}, we observe that the particle spends more time in the resetting phase for a fully asymmetric ratchet, i.e.\ $\delta = 0$ or $\delta = 1$. As this is not immediately clear from intuition alone, we explain this behaviour with the following heuristic argument. We will consider the case where the particle has had sufficient time between resetting phases to spread out uniformly across the interval as this allows us to make several simplifying assumptions about the dynamics. In this case, the probability of being found in either region upon entering the resetting phase is given by the fraction of the interval the region covers, i.e.\ $\delta$ for $i=1$ and $1 - \delta$ for $i=2$. For a sufficiently strong potential, we can further approximate the particle as being returned to the origin deterministically by the force from the ratchet. In other words, the average time spent in region $i = 1$ or region $i=2$ during the resetting phase is simply the average distance the particle travels to the origin (half the length of the region) divided by the force in that region, resulting in $\delta^2\ell^2/2\kappa$ for $i=1$ and $(1-\delta)^2\ell^2/2\kappa$ for $i=2$. Weighting these times by the probability to be found in either region, we find the average time spent in the resetting phase each cycle under these approximations is $(3\delta^2 -3\delta + 1)\ell^2/2\kappa$, which is minimised for a symmetric ratchet, $\delta = 1/2$, and maximised for a fully asymmetric ratchet, $\delta = 0$ or $\delta = 1$.

%%%%%%%%%%%%%%%%%%%%%%%%%%%%%%%%%%%%%%%%%%%%%%%%%%

% CURRENT

%%%%%%%%%%%%%%%%%%%%%%%%%%%%%%%%%%%%%%%%%%%%%%%%%%

\section{Current}\label{Sec:Current}

The currents for the diffusion $\tilde{J}_D(\tilde{x})$ and resetting $\tilde{J}_R(\tilde{x})$ phases are related to the probability densities calculated in Sec.~\ref{Sec:ProbabilityDensities} through Eqs.~(\ref{subeq:Current_DiffusionPhase}) and (\ref{subeq:Current_ResettingPhase}), respectively. The net current is then given by $\tilde{J} = \tilde{J}_D(x) + \tilde{J}_R(x)$, yielding
\begin{equation}
\begin{split}
    \tilde{J} &= \frac{p_D}{2} \frac{\kappa^2}{(e^\ell - 1)(1 - e^\kappa)\lambda_+^{[1]}\lambda_-^{[1]}\lambda_+^{[2]}\lambda_-^{[2]}} \left(\lambda_-^{[1]}\lambda_+^{[2]}e^{\lambda_+^{[1]}} + (2\delta-1)\left(\lambda_+^{[1]}\lambda_+^{[2]}e^\ell - \lambda_-^{[1]}\lambda_-^{[2]}\right) - \lambda_+^{[1]}\lambda_-^{[2]}e^{\lambda_+^{[2]}} \right),\\
\end{split}\label{eq:OverallCurrentFullExpression}
\end{equation}
where $p_D$ is fixed from either Eq.~(\ref{eq:NormalisationCondition}) or Eq.~(\ref{eq:NormalisationCondition_MFPT}), and we have reused the same compact notation $\lambda^{[1]}_{\pm} = \kappa \pm \delta\ell$, $\lambda^{[2]}_{\pm} = \kappa \pm (1 - \delta)\ell$ introduced in Eq.~(\ref{eq:ProbabilityDensity_Resetting_Solution}). In the present section, we explore the parameter space of the net current $\tilde{J}$. Given that Eq.~(\ref{eq:OverallCurrentFullExpression}) is rather opaque, we focus on exploring its more relevant limiting behaviour and general properties, as well as that of the currents associated with each phase, $\tilde{J}_D(x)$ and $\tilde{J}_R(x)$.

\begin{figure}
    \centering
    \includegraphics[width=0.7\linewidth]{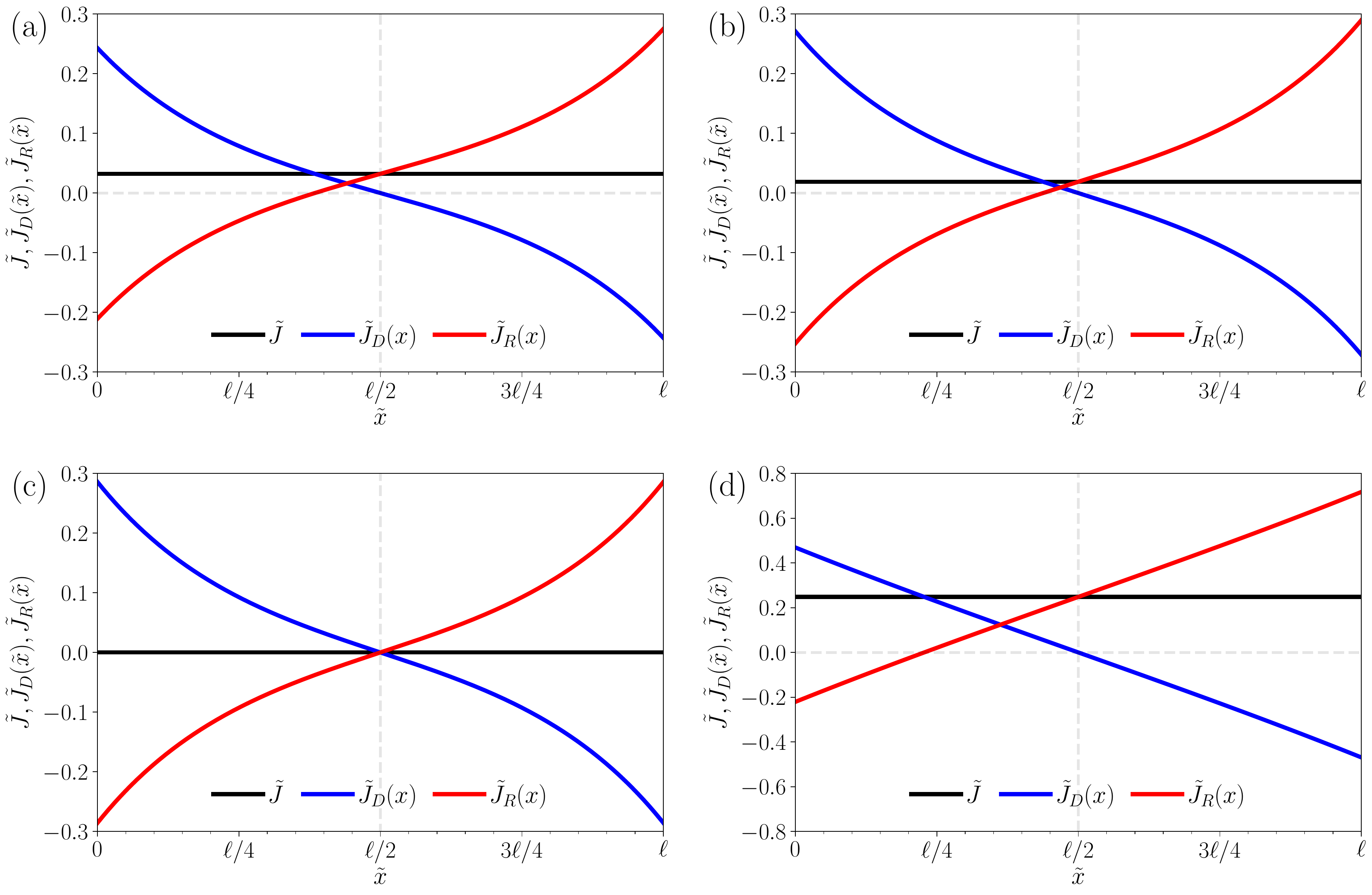}
    \caption{Steady-state currents $\tilde{J}_{D}(x)$, Eq.~(\ref{subeq:Current_DiffusionPhase}), $\tilde{J}_{R}(x)$, Eq.~(\ref{subeq:Current_ResettingPhase}), and $\tilde{J} = \tilde{J}_{D}(x) + \tilde{J}_{R}(x)$, Eq.~(\ref{eq:OverallCurrentFullExpression}), plotted for the same sets of parameters used in figure~\ref{fig:ProbabilityDensitiesVsPosition}, i.e.\ (a) $\delta = 0$, $\ell = 4$, $\kappa = 1$, (b) $\delta = 0.2$, $\ell = 4$, $\kappa = 1$, (c) $\delta = 0.5$, $\ell = 4$, $\kappa = 1$, and (d) $\delta = 0$, $\ell = 1$, $\kappa = 4$. Dashed grey lines are shown for $\tilde{J}=0$ and $\tilde{x}=\ell/2$, and act as a guide for the eye.
    }
    \label{fig:CurrentsVsPosition}
\end{figure}

The net current $\tilde{J}$, Eq.~(\ref{eq:OverallCurrentFullExpression}), and currents for each phase $\tilde{J}_{D,R}(x)$, are plotted in figure~\ref{fig:CurrentsVsPosition} for the same sets of parameters used to plot the corresponding probability densities in figure~\ref{fig:ProbabilityDensitiesVsPosition}. We observe that the diffusion-phase current is always zero at the halfway point of the interval, i.e.\ $\tilde{J}_D(\ell/2) = 0$, and that the spatially averaged diffusion-phase current vanishes, i.e.\ $\int_0^\ell d\tilde{x}~\tilde{J}_D(\tilde{x})/\ell = 0$. More generally, the diffusion-phase current is antisymmetric about the origin $\tilde{x} = 0$, i.e. $\tilde{J}_D(\tilde{x}) = -\tilde{J}_D(\ell - \tilde{x})$, where we have identified $-\tilde{x} \sim \ell - \tilde{x}$. This immediately implies the diffusion-phase current is also antisymmetric about the halfway point $\tilde{x} = \ell/2$, which can be easily seen by redefining $\tilde{x} = \tilde{x}' - \ell/2$ in the above relation. These symmetry properties arise because the diffusion phase is not biased by any potential, and so every time the particle is reset to the origin $\tilde{x} = 0$, its density proceeds to diffuse away symmetrically, see figure~\ref{fig:TimeDependentProbabilityDensities}a. Unlike the diffusion phase, the resetting-phase current $\tilde{J}_R(\tilde{x})$ is generally antisymmetric about the halfway point $\tilde{x} = \ell/2$ and the origin $\tilde{x} = 0$ \textit{only} after adjusting by a constant shift, i.e.\ $\tilde{J}_R(\tilde{x}) - \tilde{J} = -(\tilde{J}_R(\ell - \tilde{x})-\tilde{J})$. This is expected from the antisymmetry properties of $\tilde{J}_D(\tilde{x})$, since $\tilde{J}_D(\tilde{x}) = \tilde{J} - \tilde{J}_R(\tilde{x})$ by definition. As a result of this antisymmetry, the spatially averaged resetting-phase current is simply the net current, i.e.\ $\int_0^\ell d\tilde{x}~\tilde{J}_R(\tilde{x})/\ell = \tilde{J}$, indicating that the resetting-phase current is solely responsible for the non-zero net current that is generated by an asymmetric ratchet, $\delta \neq 1/2$, see figures~\ref{fig:CurrentsVsPosition}a, \ref{fig:CurrentsVsPosition}b, and \ref{fig:CurrentsVsPosition}d. On the other hand, for a symmetric ratchet, $\delta = 1/2$, the resetting-phase current has the same antisymmetry properties as the diffusion-phase current and so the net current $\tilde{J}$ vanishes, see figure~\ref{fig:CurrentsVsPosition}c. 

The role of the ratchet's symmetry is more clearly elucidated from inserting the explicit expression for the resetting-phase current $\tilde{J}_R^{[i]}(\tilde{x}) = -\partial_{\tilde{x}} \tilde{P}_R^{[i]}(\tilde{x}) - \tilde{V}^{'[i]} \tilde{P}_R^{[i]}(\tilde{x})$ into the relation $\tilde{J} = \int_0^\ell d\tilde{x}~\tilde{J}_R(\tilde{x})/\ell$ found above, whence we obtain
\begin{equation}\label{eq:OverallCurrentSymmetryRelation}
    \tilde{J} = -\frac{1}{\ell}\int_{0}^{\ell} d \tilde{x}~ \tilde{V}'(\tilde{x}) \tilde{P}_R(\tilde{x}) = \frac{\kappa}{\ell^2}\left( \frac{p_R^{[2]}}{1-\delta} - \frac{p_R^{[1]}}{\delta}\right),
\end{equation}
where we have used $\tilde{P}_R(0) = \tilde{P}_R(\ell) = 0$, Eq.~(\ref{subeq:PBC_Resetting_dimensionless}), the definitions for each $p_R^{[i]}$, Eq.~(\ref{eq:ProbabilityResettingPhase}), and have inserted the explicit expressions for each $\tilde{V}'^{[i]}$, Eq.~(\ref{eq:RatchetGradient_dimensionless}). Equation~(\ref{eq:OverallCurrentSymmetryRelation}) relates the current $\tilde{J}$ to the probabilities of being found in either region of the ratchet and shows, in particular, that the current vanishes when there is an equal chance of being found in either region, $p_R^{[1]} = p_R^{[2]}$, i.e.\ when the ratchet is symmetric $\delta = 1/2$.

\begin{figure}
    \centering
    \includegraphics[width=\linewidth, trim = 0.5cm 0.5cm 0.3cm 0cm, clip]{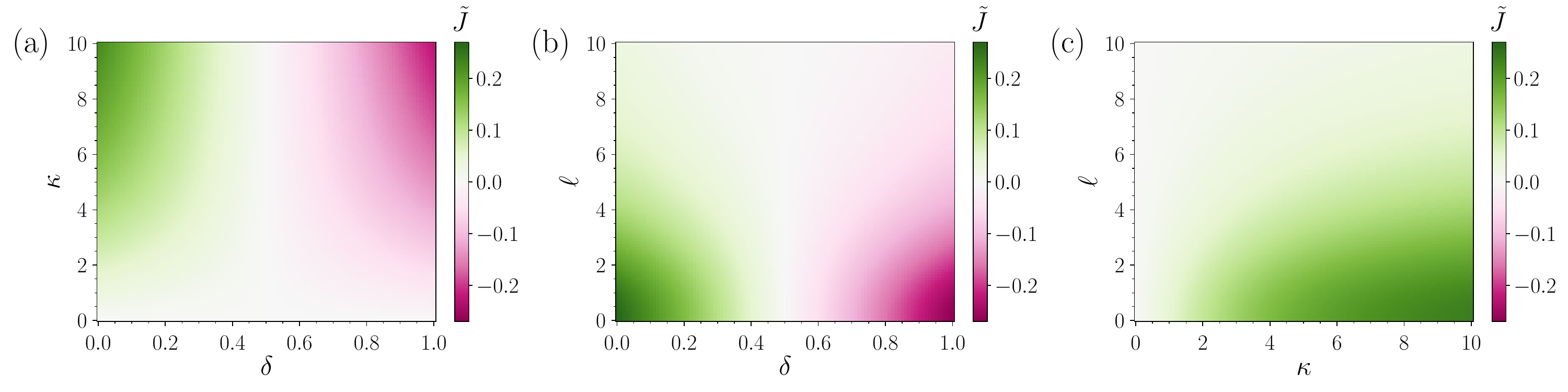}
    \caption{Steady-state current $\tilde{J}$, Eq.~(\ref{eq:OverallCurrentFullExpression}), as a function of the dimensionless parameters $\ell$, $\kappa$, and $\delta$. The constant parameter in each subfigure is (a) $\ell = 4$, (b) $\kappa = 4$, and (c) $\delta = 0.2$.}
    \label{fig:CurrentVsParameters}
\end{figure}

The net current $\tilde{J}$, Eq.~(\ref{eq:OverallCurrentFullExpression}), is plotted in figure~\ref{fig:CurrentVsParameters} as a function of the three dimensionless parameters $\ell,\kappa,\delta$. We observe that maximum positive current is obtained for a fully asymmetric ratchet $\delta = 0$,
since this removes the region of the ratchet that imparts an unfavourable negative drift. Figure~\ref{fig:CurrentVsParameters} also shows that a stronger coupling to the potential $\kappa$ leads to a larger current. This is because larger $\kappa$ reduces the time it takes to reset the particle and reduces the chance of unfavourable trajectories where the particle diffuses up the gradient of the potential. Larger $\ell$, indicating a relatively small diffusion distance $\sigma = \sqrt{D_0/r}$ between resetting phases, is seen to decrease the current. If $\ell$ satisfies either $1/\ell < \delta < 1/2$ or $1/\ell < 1-\delta < 1/2$, equivalent to $\sigma < a < L/2$ and $\sigma < L-a < L/2$, respectively, then the majority of resetting cycles will result in the particle being reset without ``completing a lap" of the system (on an extended periodic landscape, this would be equivalent to the particle being reset to the same resetting point each time). This results in an effective confinement of the particle close to the origin $\tilde{x}=0$, reducing any current generated to rare noise-induced escapes from this effective confinement.

We now consider several relevant limiting cases of the net current $\tilde{J}$, Eq.~(\ref{eq:OverallCurrentFullExpression}). First, we consider instantaneous stochastic resetting $\kappa \to \infty$. In this limit, the force from the potential is
infinitely strong and so the particle is instantaneously reset to the origin $\tilde{x} = 0$ every time the potential is switched on. We find the current in the instantaneous resetting limit $\kappa \to \infty$ is given by
\begin{equation}\label{eq:CurrentInstantaneousResetting}
    \tilde{J} \xrightarrow{\kappa \to \infty} \frac{\sinh\left(\frac{(1-2\delta)\ell}{2}\right)}{2\sinh\left(\frac{\ell}{2}\right)},
\end{equation}
which is bounded by $1/2 - \delta$ from above or below depending on whether $\delta < 1/2$ or $\delta > 1/2$, respectively, and decays monotonically to zero as $\ell \to \infty$. Equation~(\ref{eq:CurrentInstantaneousResetting}) could have equally been derived from the diffusion-phase density, Eq.~(\ref{eq:ProbabilityDensity_Diffusion_Solution}), alone using the following argument. For instantaneous resetting $\kappa\to\infty$, the net current depends solely on where the particle enters the resetting phase, as this determines the distance the particle travels upon being reset. Hence, we can deduce that the current will be given by
\begin{equation} \label{eq:CurrentInstantaneousResettingMotivation}
\tilde{J} \xrightarrow{\kappa \to \infty} \frac{1}{\ell}\int_{\delta \ell}^{\ell} dy~(\ell - y) \tilde{P}_D(y|D)  - \frac{1}{\ell}\int_{0}^{\delta \ell} dy~y\tilde{P}_D(y|D),
\end{equation}
where the first term arises from trajectories where the particle ends its diffusion phase in the $i=2$ region of the ratchet at position $y$, resulting in a jump forward by a distance $\ell - y$ upon being reset. The second term comes from the particle ending its diffusion phase in the $i=1$ region, in which case it jumps backwards a distance $y$ upon being reset. By using that $\tilde{P}_D(y|D) = \tilde{P}_D(y)$ for $\kappa \to \infty$, since $p_D = 1$ in this case, then inserting the explicit expression for $\tilde{P}_D(y)$ from Eq.~(\ref{eq:ProbabilityDensity_Diffusion_Solution}) into Eq.~(\ref{eq:CurrentInstantaneousResettingMotivation}), and finally evaluating the integrals, we recover the instantaneous resetting limit, Eq.~(\ref{eq:CurrentInstantaneousResetting}). In the limit $\kappa \to 0$, we find the current vanishes because there is no potential to rectify the particle's motion, and so its dynamics reduce to purely diffusive motion on an unbiased interval.

As illustrated in figure~\ref{fig:CurrentVsParameters}, the net current $\tilde{J}$, Eq.~(\ref{eq:OverallCurrentFullExpression}), is larger for a fully asymmetric ratchet, $\delta \to 0$, since the drift imparted by the potential is always positive in this case. In this ``ideal ratchet" limit $\delta \to 0$, we obtain the more compact expression
\begin{equation}\label{eq:CurrentIdealRatchet}
\begin{split}
    \tilde{J} &\xrightarrow{\delta \to 0}  \frac{p_D}{2} \frac{ \kappa^2(e^{\ell} - 1)(e^{\kappa} +1) - \kappa \ell (e^{\ell} + 1)(e^{\kappa} - 1) }{ (\kappa^2 - \ell^2)(e^\kappa -1)(e^\ell -1) }\\ &= \frac{ \kappa^2(e^{\ell} - 1)(e^{\kappa} +1) - \kappa \ell (e^{\ell} + 1)(e^{\kappa} - 1) }{2\kappa^2(e^{\kappa} - 1)(e^{\ell} - 1) + \kappa\ell^2(e^\kappa + 1)(e^\ell - 1) - \ell^2(e^\kappa - 1)(\ell - 2 + e^\ell(\ell+2))} ,
\end{split}
\end{equation}
while the opposite limit $\delta \to 1$ simply returns the above expression with an overall minus sign.

Figure~\ref{fig:CurrentVsParameters}a suggests the maximum possible current is obtained for a ratchet that is both fully asymmetric $\delta \to 0$, and infinitely strong $\kappa \to \infty$. This corresponds to an instantaneous resetting model in which a particle jumps forward every resetting event. Taking both $\kappa \to \infty$ and $\delta \to 0$ in Eq.~(\ref{eq:OverallCurrentFullExpression}), we find $\tilde{J} \to 1/2$, which also establishes an upper bound on the current that can be generated. This could have equally been derived through the following physical argument. Suppose the diffusion phase ends with the particle located a distance $y$ from the origin $\tilde{x} = 0$. Then, upon being reset, the particle will jump forward by either $y$ or $\ell - y$, i.e.\ an average distance $\ell/2$. Given the diffusion-phase dynamics are symmetric and hence do not contribute to the net current, the average number of times the particle passes a point in the interval of length $\ell$ is therefore $\tilde{J} = 1/2$. In dimensionful units, this equates to $J = r/2$, and so the physical current can be made arbitrarily large by increasing the resetting rate, since this corresponds to an infinitely frequent series of instantaneous jumps forward.

\begin{figure}
    \centering
    \includegraphics[width=0.7\linewidth, trim = 0.7cm 0cm 16cm 0cm, clip]{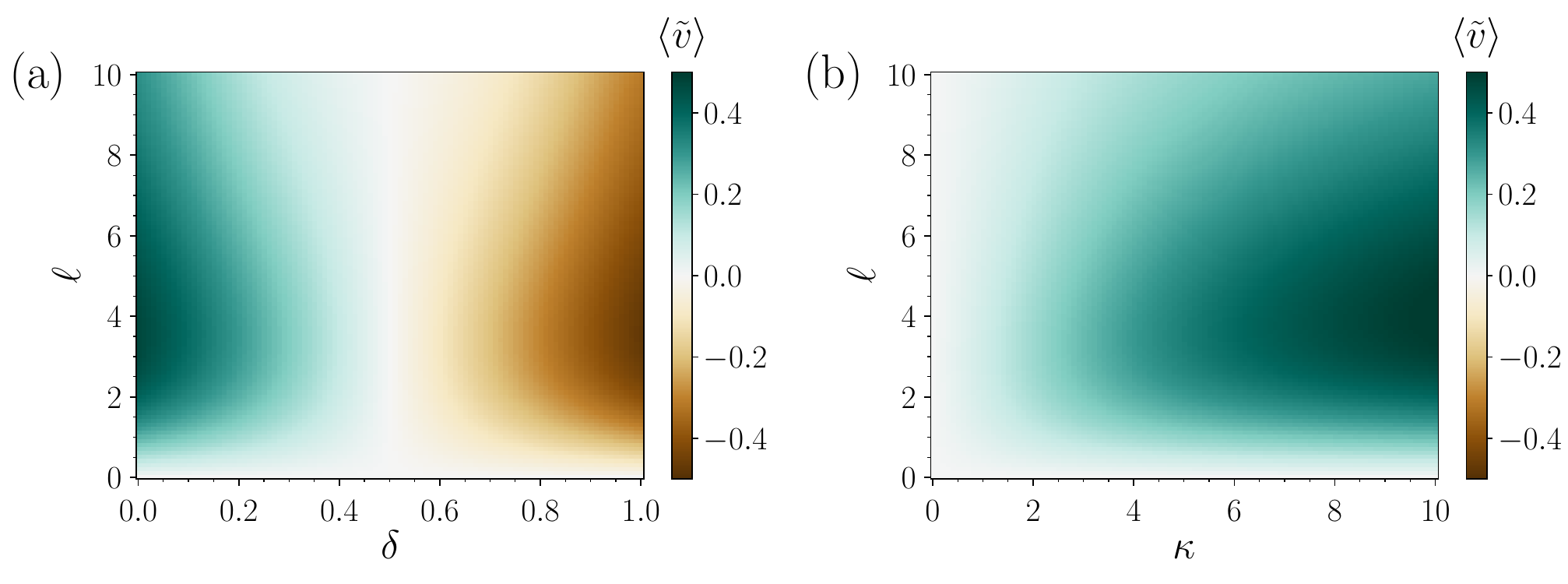}
    \caption{Steady-state average drift speed $\langle \tilde{v} \rangle$, Eq.~(\ref{eq:ParticleAverageSpeed}), as a function of the dimensionless parameters $\ell$, $\kappa$, and $\delta$. The constant parameter in each subfigure is (a) $\kappa = 4$ and (b) $\delta = 0.2$.}
    \label{fig:VelocityVsParameters}
\end{figure}

As seen in figure~\ref{fig:CurrentVsParameters}, the current's dependence on the system parameters is monotonic. However, the average drift speed of the particle's rectified motion, given by
\begin{equation}
\label{eq:ParticleAverageSpeed}
    \langle \tilde{v} \rangle = \tilde{J} \ell
\end{equation}
in dimensionless units, has non-monotonic dependence on the system length $\ell$ for finite $\kappa$ or $\delta$, as shown in figure~\ref{fig:VelocityVsParameters}. For $\kappa \to \infty$ and $\delta \to 0$, however, we find $\langle \tilde{v} \rangle \to \ell/2$, suggesting the particle's drift speed diverges as $\ell \to \infty$. In this case, it is advantageous to have a larger system length $\ell$ for obtaining greater drift speed, as the particle always jumps forward upon being reset and the distance it jumps is greater for larger $\ell$.

%%%%%%%%%%%%%%%%%%%%%%%%%%%%%%%%%%%%%%%%%%%%%%%%%%%

% Efficiency

%%%%%%%%%%%%%%%%%%%%%%%%%%%%%%%%%%%%%%%%%%%%%%

\section{Energetic cost and efficiency of resetting}\label{sec:Efficiency}

In the previous section, we found the dimensionless current $\tilde{J}$ is trivially maximised by taking $\delta \to 0$ and $\kappa \to \infty$, the latter being equivalent to an infinitely strong potential. The physical current $J = r\tilde{J}$ is then maximised by an infinite resetting rate $r \to \infty$. While maximising the current is a physically relevant objective, it does not take into account the cost associated with using the potential to reset the particle. In particular, we expect a potential of greater amplitude to incur a greater energetic cost. A larger resetting rate $r$ would similarly represent a larger rate with which energy needs to be supplied to the system.

To quantify the energetic cost associated with producing a current in this system, we are motivated to study a suitable efficiency parameter. For non-equilibrium systems, the thermodynamic efficiency $\eta_{\mathrm{th}} = \dot{W}/(\dot{W} + D_0\dot{S})$ is typically studied \cite{pietzonka2019, roberts2023}, where $\dot{W}$ is the useful power that can be extracted from the system (such as by doing work against an external force \cite{pietzonka2019,roberts2023,cocconi2023,cocconi2024,derivaux2023}) and $\dot{S}$ is the entropy production rate \cite{cocconi2020}, which quantifies the irreversibility of the dynamics. As for many resetting models, the entropy production of the present system diverges due to the resetting process being unidirectional \cite{pal2021}. Hence, the thermodynamic efficiency vanishes and therefore cannot be used to distinguish between different parameters. This is avoided in approximate resetting schemes, such as motion in a fluctuating confining potential \cite{alston2022, olsen2023}, or resetting to a distribution \cite{bressloff2024, mori2023}, for which the corresponding entropy production rates are finite. However, for single-point resetting, an alternative way of quantifying the energetic cost must be sought.

\begin{figure}
    \centering
    \includegraphics[width=\linewidth]{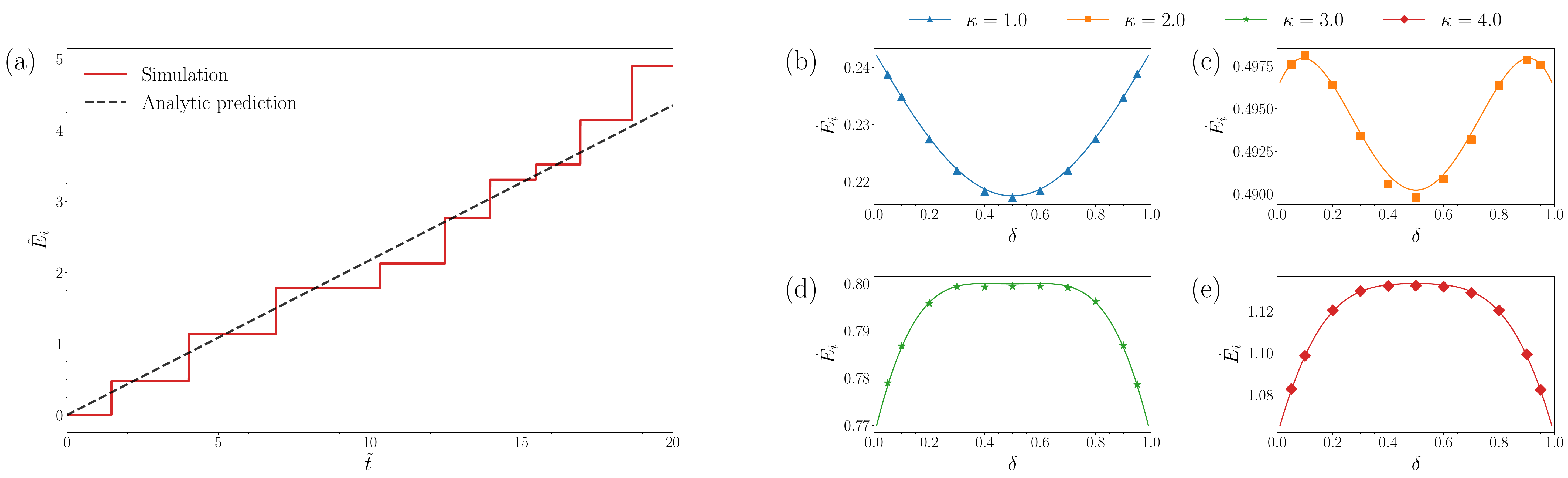}
    \caption{Energetic cost of resetting compared to simulations. (a) Cumulative energy input for a single simulated particle against time. The analytical prediction for the cumulative energy input is obtained by multiplying the exact analytic expression for energy input rate, Eq.~(\ref{eq:energeticcostexpr_exact}), by the elapsed time, and is plotted as a dashed line. In the simulation, the cumulative energy input experiences discontinuous jumps of $\tilde{V}(\tilde{x})$ each time the potential is switched on, i.e.\ when the particle goes from the diffusion phase to the resetting phase. The simulation parameters are $\ell=4$, $\kappa=1$, $\delta=0.5$.
    (b)-(e) Rate of energy input $\dot{E}_i$, Eq.~(\ref{eq:energeticcostdefn}), compared to simulations. The theoretical results from Eq.~(\ref{eq:energeticcostexpr_exact}) are shown as solid lines. The simulation data are plotted as symbols for $\kappa=1,2,3,4$, against various values of $\delta \in (0,1)$, with $\ell=4$. The simulations show very good agreement with the analytical results. The simulated energy input rates were calculated by averaging $10^4$ particle realisations up to a total time $\tilde{t}=500$. The simulations of the energetic cost were performed with a smaller timestep $\delta \tilde{t} = 10^{-5}$ than the other simulations in this work due to the small variation in energetic cost (see, in particular, subfigure (c)) necessitating a higher degree of accuracy.}
    \label{fig:CostFunctionVsSimulations}
\end{figure}

One recent proposal for an appropriate cost function associated with resetting is based on the distance the particle travels upon being reset \cite{sunil2023}. A drawback of this is that it does not take into account the nature of the mechanism by which the particle is reset. A different proposal that is more grounded in thermodynamics is based on an experimental implementation of stochastic resetting, where the cost of a reset is defined as the total energy supplied to an experimental trapping device, such as an optical trap, over the duration of the resetting phase \cite{tal-friedman2020}. In the same vein, we define an experimentally motivated, yet analytically tractable, cost function given by the average rate of work done by the potential in order to repeatedly return the particle to the origin. The work done over the course of a single resetting phase is equal to the difference in potential energy of the particle at the beginning and end of the phase \cite{alston2022, olsen2023, gupta2022a}. Since every stochastic trajectory in the resetting phase ends at $x=0$, where $V(0) = 0$, then the net work done in a resetting phase starting at position $x = x_s$ is simply $V(x_s)$. Said another way, the average rate of work done is equal to the power supplied to the system to sustain the repeated jumps in the particle's potential energy from switching on the resetting potential, see figure~\ref{fig:CostFunctionVsSimulations}a. 
Therefore, the average work done over a complete cycle of diffusion and resetting phases is $\int_0^L dx~V(x)P(x|D)$, where $P(x|D) = P_D(x)/p_D$ is the conditional probability density for the particle to be found at position $x$ \textit{given} it is in the diffusion phase. Meanwhile, the average time for a resetting cycle is $\tau + \int_0^L dx~T_R(x)P(x|D) = 1/(rp_D)$, see Eq.~(\ref{eq:NormalisationCondition_MFPT}). Hence, by dividing the average work done over a cycle by the average time for a cycle, we obtain for the steady-state rate of energy input/work done,
\begin{equation}
    \label{eq:energeticcostdefn}
    \dot{E}_{i} = r \int_{0}^{L} dx~V(x)P_D(x).
\end{equation}
Equation~(\ref{eq:energeticcostdefn}) vanishes for either $h \to 0$ or $r \to 0$, and diverges in the instantaneous resetting limit $h \to \infty$,
as we would expect for a physically meaningful measure of the energetic cost associated with resetting.

\begin{figure}
    \centering
    \includegraphics[width=\linewidth, trim = 0.4cm 0.5cm 0.1cm 0.2cm, clip]{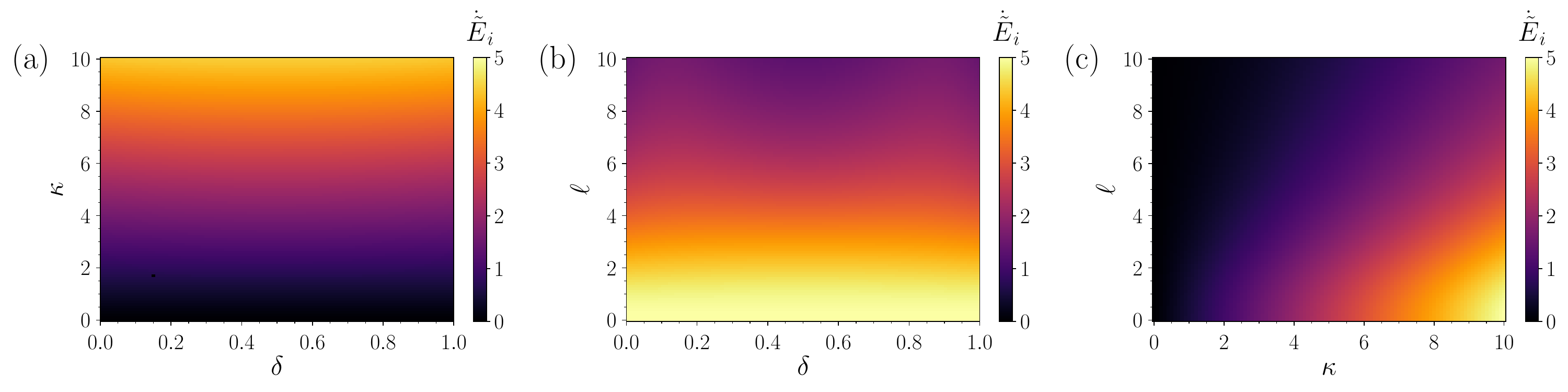}
    \caption{Steady-state energy input rate $\dot{\tilde{E}}_i$, Eq.~(\ref{eq:energeticcostexpr_exact}), as a function of the dimensionless parameters $\ell$, $\kappa$, and $\delta$. The constant parameter in each subfigure is (a) $\ell = 2$, (b) $\kappa = 10$, and (c) $\delta = 0.2$.
    }
    \label{fig:CostFunctionVsParameters}
\end{figure}

By evaluating the integral in Eq.~(\ref{eq:energeticcostdefn}), we obtain the following expression for the energy input rate in terms of the dimensionless parameters and $p_D$,
\begin{equation}
    \label{eq:energeticcostexpr_exact}
    \dot{\tilde{E}}_{i} = p_D \frac{ \kappa\sinh\left(\frac{\delta\ell}{2}\right) \sinh\left(\frac{(1-\delta)\ell}{2}\right)}{\delta \ell (1-\delta)\sinh\left(\frac{\ell}{2}\right)}.
\end{equation}
In figure~\ref{fig:CostFunctionVsSimulations}, we show that Eq.~(\ref{eq:energeticcostexpr_exact}) is in good agreement with the simulation results. In figure~\ref{fig:CostFunctionVsParameters}, we observe that the energy input rate generally increases with $\kappa$ and decreases with $\ell$. The energy input rate decreases with $\ell$ because larger $\ell$ represents a relatively smaller diffusion distance $\sigma = \sqrt{D_0/r}$ between resets, meaning the particle is effectively confined closer to the origin $\tilde{x} = 0$ between resets, and so the potential does relatively little work to reset the particle each cycle. Varying the asymmetry parameter $\delta$ has comparatively little effect on the energy input rate, but there is an interesting non-monotonic dependence on $\delta$ nonetheless, as can be seen in figures~\ref{fig:CostFunctionVsSimulations}b-e.

In the fully asymmetric limit $\delta \to 0$, the energy input rate can be written out exactly in terms of the dimensionless parameters, i.e.
\begin{equation}
    \label{eq:energeticcostexpr_limit}
    \dot{\tilde{E}}_{i} \xrightarrow{\delta \to 0} p_D\frac{\kappa}{2} = \frac{\kappa(\kappa^2 - \ell^2)(e^{\kappa} - 1)(e^{\ell} - 1)}{2\kappa^2(e^{\kappa} - 1)(e^{\ell} - 1) + \kappa\ell^2(e^\kappa + 1)(e^\ell - 1) - \ell^2(e^\kappa - 1)(\ell - 2 + e^\ell(\ell+2))},
\end{equation}
and so, in this case, the rate that energy is supplied to the system is simply equal to the average height of the potential, $\kappa/2$, divided by the total time for a complete cycle, $1/p_D$, in dimensionless units.

\begin{figure}
    \centering
    \includegraphics[width=\textwidth]{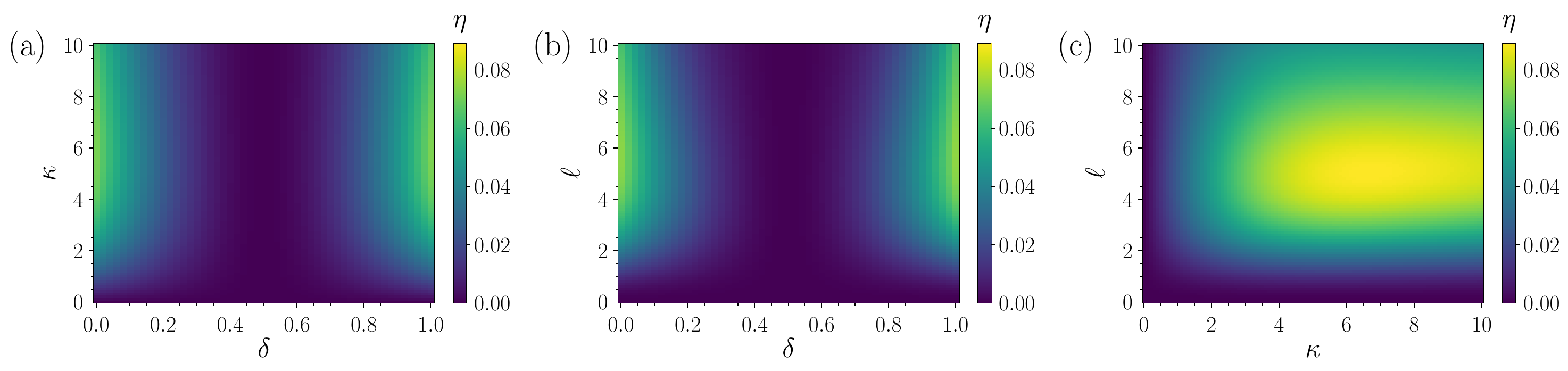}
    \caption{Efficiency $\eta$, Eq.~(\ref{eq:efficiency_explicit}), as a function of the dimensionless parameters $\ell$, $\kappa$, and $\delta$. The constant parameter in each subfigure is (a) $\ell = 4$, (b) $\kappa = 4$, and (c) $\delta = 0.2$. The maximum efficiency in each subfigure is achieved at (a) $\delta=0$, $\kappa=5.9$; (b) $\delta=0$, $\ell=5.7$; and (c) $\kappa=6.7$, $\ell=5.1$.}
    \label{fig:efficiency}
\end{figure}

\begin{figure}
    \centering
    \includegraphics[width=0.5\textwidth]{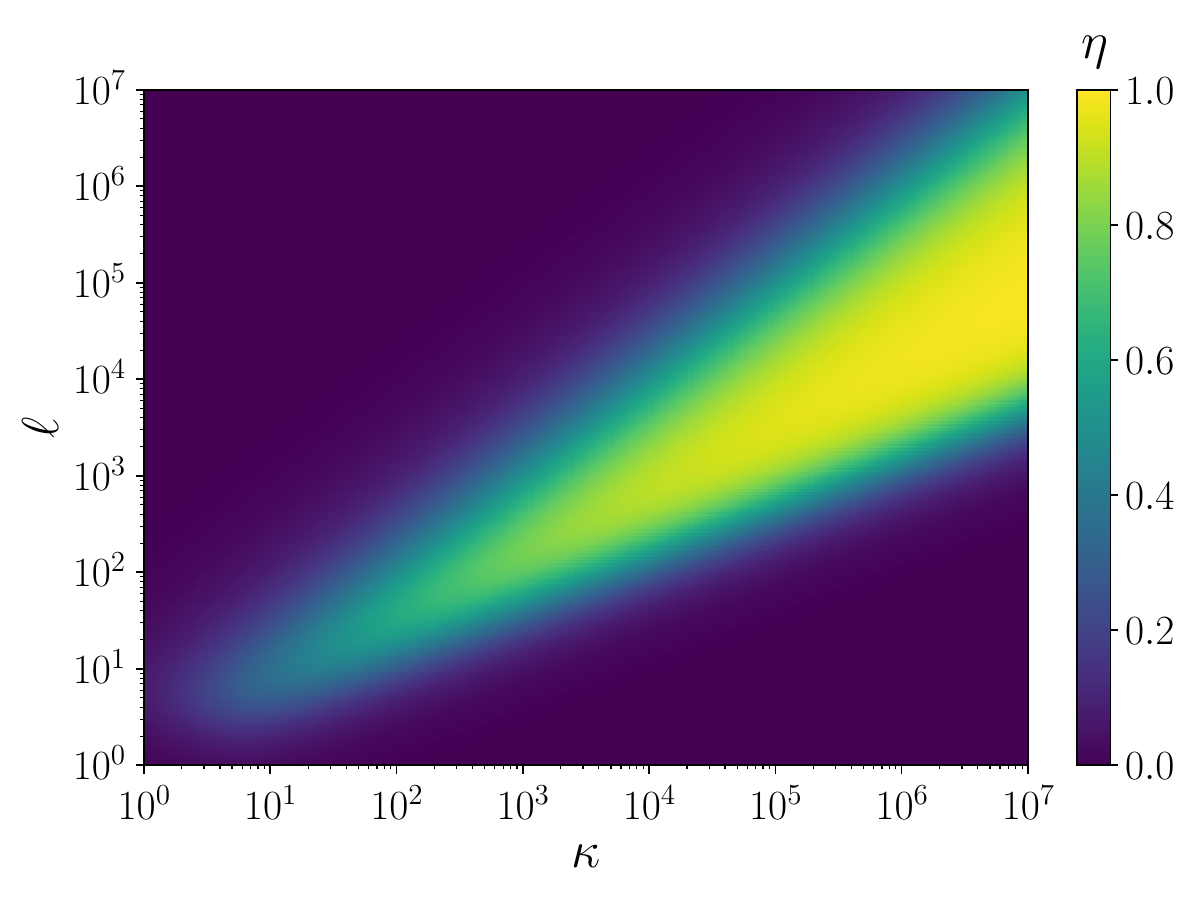}
    \caption{Efficiency $\eta$ in the limit $\delta\to 0$, Eq.~(\ref{eq:efficiency_limit}), as a function of $\kappa$ and $\ell$ on a log-log scale.}
    \label{fig:efficiency_limit}
\end{figure}

We now wish to compare the energetic cost of resetting against the return we get in current, similarly to the cost-weighted efficacy \cite{sunil2024} and payoff \cite{debruyne2023} functions recently defined for diffusive search processes with stochastic resetting. We define our efficiency parameter as
\begin{equation} \label{eq:efficiency_definition}
    \eta = \frac{\langle \tilde{v} \rangle^2}{ \dot{\tilde{E}}_{i}},
\end{equation}
where, in our units in which the friction coefficient has been set to unity, $\langle \tilde{v} \rangle^2$ is the energy per unit time associated with the current flow. Thus, $\eta$, which is bounded by $0$ and $1$, can be thought of as the ratio of useful power output $\langle v \rangle^2$ to power input $\dot{E}_i$. The efficiency can be expressed compactly in terms of the probabilities to be in either phase and the dimensionless parameters,
\begin{equation} \label{eq:efficiency_explicit}
    \eta = \frac{1 }{ p_D} \left(\frac{p_R^{[2]}}{1-\delta} - \frac{p_R^{[1]}}{\delta}\right)^2\frac{ \kappa \delta (1- \delta) \sinh\left( \frac{\ell}{2}\right)}{\ell \sinh\left( \frac{\delta \ell}{2}\right) \sinh\left( \frac{(1 - \delta) \ell}{2}\right)},
\end{equation}
where we stress again that $p_D$ and $p_R$ have closed-form expressions, determined from Eq.~(\ref{eq:NormalisationCondition}), but they are too complicated to write out explicitly here. Equation (\ref{eq:efficiency_explicit}) is plotted in figure~\ref{fig:efficiency}. We observe that the efficiency has a non-monotonic dependence on $\kappa$ and $\ell$ due to a compromise between larger $\kappa$ and smaller $\ell$ resulting in greater current, but also incurring a greater energetic cost. The efficiency is generally greater for a fully asymmetric ratchet, $\delta \to 0$ or $\delta \to 1$, due to the magnitude of the particle's average drift speed $\langle \tilde{v} \rangle$ increasing faster than the energetic cost $\dot{\tilde{E}}_i$ as $\delta \to 0$ or $\delta \to 1$. In the limit $\delta \to 0$, the efficiency reduces to the exact expression,
\begin{equation}
    \label{eq:efficiency_limit}
    \eta \xrightarrow{\delta\to0} \frac{\kappa \ell^2 (\ell \coth(\ell/2) - \kappa \coth(\kappa/2))^2}{(\ell^2 - \kappa^2)(2(\ell^2 - \kappa^2) + \ell^3 \coth(\ell/2) -\ell^2 \kappa \coth(\kappa/2))},
\end{equation}
which is plotted in figure~\ref{fig:efficiency_limit}, and could have equally been obtained by taking $\delta \to 1$ in Eq.~(\ref{eq:efficiency_explicit}). One can now find values for the remaining parameters to optimise the efficiency. For instance, the optimal $\kappa = \kappa^*$ for a given $\ell$ that maximises the efficiency is found by solving $\left. \partial_\kappa \eta \right|_{\kappa = \kappa*} = 0$.

Figure~\ref{fig:efficiency_limit} demonstrates that arbitrarily perfect efficiency can be achieved for a region of parameter space where both $\kappa \to \infty$ and $\ell \to \infty$. However, na\"{i}vely taking $\kappa \to \infty$ or $\ell \to \infty$ in Eq.~(\ref{eq:efficiency_limit}) returns zero efficiency. Extra care needs to be taken when taking such limits. Physically, we expect $D_0 \rightarrow 0$ to be optimal as it removes the possibility of unfavourable trajectories where the particle diffuses up the potential gradient. To take $D_0 \to 0$, we can take $\kappa \to \infty$ and $\ell \to \infty$ while keeping $\kappa/\ell^2 = c = h/(L^2 r)$ a finite constant. This yields a finite efficiency $\eta = 1/(1 + 2c)$, implying that decreasing $c$ will result in a better efficiency. One may then be tempted to set $c = 0$ to yield perfect efficiency. However, this breaks our previous requirement of $c$ being a finite constant. Hence, extra care needs to be taken when taking these multiple limits. In particular, one needs to ensure $D_0 \to 0$ much faster than $c \to 0$. These limits are, of course, purely mathematical, but give useful insight into how one could optimise the efficiency of potential-mediated resetting nonetheless.

%%%%%%%%%%%%%%%%%%%%%%%%%%%%%%%%%%%%%%%%%%%%%%%%%%

% CONCLUSION

%%%%%%%%%%%%%%%%%%%%%%%%%%%%%%%%%%%%%%%%%%%%%%%%%%

\section{Conclusion}\label{Sec:Conclusion}

In this paper, we considered an experimentally motivated resetting model, whereby the resetting of a particle to its initial position is mediated by a periodic ratchet potential, Eq.~(\ref{eq:ratchetpotential}), allowing a steady-state net current to be generated. In particular, the ratchet potential switches on with a constant rate, but switches off only upon the particle reaching the minimum of the potential. Not only does this constitute an optimised variation of the fluctuating potential ratchet model \cite{reimann2002}, it also generalises other well-known resetting models \cite{gupta2021, evans2011}, which can be recovered by taking certain limits.

We derived the exact solutions for the steady-state phase-dependent probability densities, Eqs.~(\ref{eq:ProbabilityDensity_Diffusion_UnspecifiedConstants}) and (\ref{eq:ProbabilityDensity_Resetting_Solution}), which were analysed in detail in Sec.~\ref{Sec:ProbabilityDensities}, and corroborated these with particle-based Monte Carlo simulations, see figures~\ref{fig:TimeDependentProbabilityDensities} and \ref{fig:ProbabilityDensitiesVsPosition}. This was supplemented by a derivation of the time-dependent densities in Appendix~\ref{App:Sec:DensitySolutions}, as well as a derivation of the steady-state densities for a similar ``stop-and-start" resetting model in Appendix~\ref{App:Sec:DeterministicRatchetModel}. From the densities we derived an exact expression for the steady-state net current, Eq.~(\ref{eq:OverallCurrentFullExpression}), in Sec.~\ref{Sec:Current}. We showed the net current is tightly bound from above by $r/2$, where $r$ is the rate the potential is switched on, by considering the optimal limits of an infinitely strong, Eq.~(\ref{eq:CurrentInstantaneousResetting}), and fully asymmetric, Eq.~(\ref{eq:CurrentIdealRatchet}), potential. As far as we are aware, this is the first analytic study of a steady-state net current generated through finite-time resetting. We made connection with experiments and work extraction in Sec.~\ref{sec:Efficiency}, where we calculated the power input that would be required to sustain this system, Eq.~(\ref{eq:energeticcostdefn}), and compared it with simulations, see figure~\ref{fig:CostFunctionVsSimulations}. We then went further by defining an efficiency parameter that quantifies the useful power output in the form of current for a given power input, Eq.~(\ref{eq:efficiency_definition}). We showed that this efficiency parameter is bounded between $0$ and $1$, before finding the parameters that yield arbitrarily perfect efficiency.

We believe this work can be extended in several ways. Firstly, we formulated the model for overdamped dynamics, and so it could be generalised to underdamped dynamics by incorporating inertia. In particular, it would be interesting to see in what parameter regimes inertia improves the current. One may expect inertia to enhance the current for a fully asymmetric ratchet where the force from the ratchet is always in the desired direction, but hindered for finite apex positions as then the ratchet has a region of much larger force in the reverse direction. Another possible avenue for further research is to allow the resetting phase to end at a position determined by a distribution, such as a Gaussian of finite width centred at the minimum of the potential. Allowing the resetting phase to potentially end at any position renders the entropy production finite, which would then pave the way for a study of the thermodynamic efficiency. An alternative approach could be to associate some inherent precision error with determining when the particle has reached the potential minimum in the resetting phase. As the error is made smaller, the associated cost of measurement goes up \cite{cocconi2024}, and so this becomes an additional consideration for finite-time resetting schemes where the resetting potential is modulated by a hypothetical external controller. If we consider the potential to be periodic on an extended domain, then the present model becomes exactly a finite-time resetting variation of the local resetting model studied in Ref.~\cite{alston2024}. One could then investigate how the addition of a finite resetting time affects the boost in late-time diffusion seen in this local resetting model. Finally, an interesting analogue to the stop-and-start model we considered in Appendix~\ref{App:Sec:DeterministicRatchetModel} could be an active particle whose self-propulsion stochastically switches off with a constant rate, and switches back on again only upon the particle reaching the potential minimum. Active particles in static ratchets generally prefer to move up the shallower slope of the potential \cite{roberts2023, zhen2022}, whereas the stop-and-start resetting scheme tends to rectify particles down the shallower slope. An active stop-and-start resetting model may then have non-monotonic dependence on the system parameters to reflect this competition, as well as the possibility of displaying current reversals.

%%%%%%%%%%%%%%%%%%%%%%%%%%%%%%%%%%%%%%%%%%%%%%%%%%

% ACKNOWLEDGEMENTS

%%%%%%%%%%%%%%%%%%%%%%%%%%%%%%%%%%%%%%%%%%%%%%%%%%

\begin{acknowledgments}
The authors thank Henry Alston, Thibault Bertrand, Gunnar Pruessner, and Letian Chen for useful discussions. C.R.\ acknowledges support from the Engineering and Physical Sciences Research Council (Grant No.\ 2478322). E.L.\ was funded by the President’s
PhD Scholarship at Imperial College London. E.S.\ was funded by the Roth Scholarship at Imperial College London.
\end{acknowledgments}

%%%%%%%%%%%%%%%%%%%%%%%%%%%%%%%%%%%%%%%%%%%%%%%%%%

% APPENDIX

%%%%%%%%%%%%%%%%%%%%%%%%%%%%%%%%%%%%%%%%%%%%%%%%%%

\appendix

\section{Simulation details} \label{App:Sec:Simulations}

By rescaling space and time by $\sigma=\sqrt{D_0/r}$ and $\tau=1/r$, respectively, we can rewrite the Langevin equations (\ref{eq:LangevinEquationDiffusion})-(\ref{eq:LangevinEquationResetting}) in terms of the nondimensional parameters $\ell, \kappa, \delta$ defined in the main text, i.e.\
\begin{subequations}\label{eq:DimensionlessLangevin_appendix}
    \begin{align} 
    \dot{\tilde{x}}_D (\tilde{t}) &= \sqrt{2}\xi(\tilde{t}), \label{eq:LangevinEquationDiffusionND} \\  
    \dot{\tilde{x}}_R(\tilde{t}) &= -\tilde{V}'\left(\tilde{x}_R(\tilde{t})\right) + \sqrt{2}\xi(\tilde{t}), \label{eq:LangevinEquationResettingND}
\end{align}
\end{subequations}
where the gradient of the ratchet potential $\tilde{V}'(\tilde{x})$ is given by Eq.~(\ref{eq:RatchetGradient_dimensionless}). The probability at each timestep to go from the diffusion phase (\ref{eq:LangevinEquationDiffusionND}) to the resetting phase (\ref{eq:LangevinEquationResettingND}) is $\Delta \tilde{t}$, i.e.\ the timestep duration,
 and the particle returns to the diffusion phase from the resetting phase with probability 1 upon crossing either $\tilde{x}=0$ or $\tilde{x}=\ell$. We solve Eq.~(\ref{eq:DimensionlessLangevin_appendix}) numerically using the Euler-Maruyama method with timestep $\Delta \tilde{t}=10^{-4}$. Simulations were run up till a final time $\tilde{t}=20$ for the numerical estimates of the steady-state probability densities in figure~\ref{fig:ProbabilityDensitiesVsPosition}, and were averaged over $10^6$ realisations.

%%%%%%%%%%%%%%%%%%%%%%%%%%%%%%%%%%%%%%%%%%%%%%%%%%%%%%%%%

\section{Time-dependent solutions for probability densities} \label{App:Sec:DensitySolutions}
In this appendix, we demonstrate how to derive the time-dependent probability densities $P_{D,R}(x,t)$. For clarity, we will perform the derivation in terms of the dimensionful variables $h$, $a$, $L$, $D_0$, and $r$, whose definitions can be found in the main text. An analogous derivation for a different model appears in Ref.~\cite{gupta2021}. The time-dependent probability densities obey the coupled Fokker-Planck equations (\ref{eq:FPequation}) which, for convenience, we repeat here:
\begin{subequations}
    \label{eq:FP_appendix}
    \begin{align}
         \frac{\partial P_D(x,t)}{\partial t} &= D_0\frac{\partial^2 P_D(x,t)}{\partial x^2} - r P_D(x,t) + \left(J_R^{[2]}(L^{-},t) - J_R^{[1]}(0^{+},t)\right) \delta(x) ,\label{subeq:FP_diffusion_appendix} \\
         \frac{\partial P_R^{[1]}(x,t)}{\partial t} &= D_0\frac{\partial^2 P_R^{[1]}(x,t)}{\partial x^2}  + V'^{[1]}\frac{\partial P_R^{[1]}(x,t)}{\partial x} + r P_D(x) + J_R^{[1]}(0^{+},t)\delta(x),
         \label{subeq:FP_resetting_appendix1}\\
         \frac{\partial P_R^{[2]}(x,t)}{\partial t} &= D_0\frac{\partial^2 P_R^{[2]}(x,t)}{\partial x^2}  + V'^{[2]}\frac{\partial P_R^{[2]}(x,t)}{\partial x} + r P_D(x) - J_R^{[2]}(L^{-},t)\delta(x).
         \label{subeq:FP_resetting_appendix2}
    \end{align}
\end{subequations}
As in the main text, we separate the solution for the resetting-phase density into two separate components
\begin{equation}\label{eq:Densities_Separate_appendix}
    P_R(x,t) = 
    \begin{cases}
      P_R^{[1]}(x,t), & 0 \leq x < a, \\
      P_R^{[2]}(x,t), & a \leq x < L,
    \end{cases}
\end{equation}
to align with the different regions of the potential, which has gradients given by
\begin{equation}
\label{eq:RatchetGradient_appendix}
    V'(x) = 
    \begin{cases}
     V'^{[1]} = \frac{h}{a}, & 0 \leq x <  a, \\
      V'^{[2]} = -\frac{h}{L-a}, &  a \leq x < L.
    \end{cases}
\end{equation}
We begin with the time-dependent solution for the diffusion-phase probability density $P_D(x,t)$, described by Eq.~(\ref{subeq:FP_diffusion_appendix}). In addition to the periodic boundary conditions,
\begin{subequations}
\label{eq:PBCsAppendix}
\begin{align}
    P_D(0,t) &= P_D(L,t), \label{eq:PBCappendixDiffusion}\\
    P_R(0,t) &= P_R(L,t) = 0, \label{eq:PBCappendixResetting}
\end{align}
\end{subequations}
we define the initial conditions as
\begin{subequations}
\label{eq:InitialConditionsAppendix}
\begin{align}
    P_D(x,0) &= \delta(x), \label{eq:InitialConditionAppendixDiffusion}\\
    P_R(x,0) &= 0, \label{eq:InitialConditionAppendixResetting}
\end{align}
\end{subequations}
i.e.\ we assume the particle is initialised at the origin in the diffusion phase. From the definition of the Laplace transform,
\begin{equation}
\label{eq:LaplaceTransformDefinition}
    \hat{P}_{D,R}(x,s) = \int_0^\infty dt~e^{-st}P_{D,R}(x,t),
\end{equation}
we obtain for the Laplace-transformed diffusion-phase density,
\begin{equation}
\label{eq:Laplacepde1}
    D_0 \frac{\partial^2 \hat{P}_D(x,s)}{\partial x^2} - (s+r)\hat{P}_D(x,s) + \left( 1+ \hat{J}_R^{[2]}(L^{-},s) - \hat{J}_R^{[1]}(0^{+},s) \right)\delta(x) = 0,
\end{equation}
where we have used that
\begin{equation}
\label{eq:LaplaceTimeDerivative}
    \int_0^\infty dt~e^{-st}\frac{\partial P_D(x,t)}{\partial t} = s \hat{P}_D(x,s) - \delta(x)
\end{equation}
after integrating by parts, recalling the initial condition $P_D(x,0) = \delta(x)$, Eq.~(\ref{eq:InitialConditionAppendixDiffusion}).

Integrating Eq.~(\ref{eq:Laplacepde1}) over the region $[-\epsilon, \epsilon]$, recalling $-\epsilon \sim L-\epsilon$, and taking $\epsilon \to 0^+$, we obtain the following jump condition at the origin,
\begin{equation}
\label{eq:jumpcondition_D}
    D_0\left( \left. \frac{\partial \hat{P}_D(x,s)}{\partial x}\right|_{x \to 0^{+}} - \left. \frac{\partial \hat{P}_D(x,s)}{\partial x}\right|_{x \to L^{-}}\right) = -\left( 1+ \hat{J}_R^{[2]}(L^{-},s) - \hat{J}_R^{[1]}(0^{+},s) \right),
\end{equation}
where $\hat{J}_R^{[2]}(L^{-},s) - \hat{J}_R^{[1]}(0^{+},s)$ will be fixed in terms of the parameters later by Eq.~(\ref{eq:JumpConditionResettingOrigin_Appendix}).

We now consider Eq.~(\ref{eq:Laplacepde1}) in the region $x \in (0,L)$, where the term proportional to the Dirac delta function vanishes. Solving in this region, we obtain
\begin{equation}\label{eq:DiffusionSolution_Appendix}
    \hat{P}_D(x,s) = \hat{B}_D(s)e^{\sqrt{\frac{s+r}{D_0}} x} + \hat{C}_D(s)e^{-\sqrt{\frac{s+r}{D_0}} x},
\end{equation}
where $\hat{B}_D(s)$ and $\hat{C}_D(s)$ are constants to be determined. The periodic boundary condition (\ref{eq:PBCappendixDiffusion}) implies
\begin{equation}\label{eq:ConstantsRelationAppendix}
   \hat{C}_D(s) = \frac{\exp\left(\sqrt{\frac{s+r}{D_0}} L\right) - 1}{1 - \exp\left(-\sqrt{\frac{s+r}{D_0}} L\right)}\hat{B}_D(s).
\end{equation}
Then, inserting the expression for $\hat{C}_D(s)$, Eq.~(\ref{eq:ConstantsRelationAppendix}), into Eq.~(\ref{eq:DiffusionSolution_Appendix}) and applying the jump condition (\ref{eq:jumpcondition_D}), we obtain
\begin{subequations}
\begin{align}
    \hat{B}_D(s) &= \frac{1+ \hat{J}_R^{[2]}(L^{-},s) - \hat{J}_R^{[1]}(0^{+},s)}{2\sqrt{D_0(r+s)}\left(\exp\left(\sqrt{\frac{s+r}{D_0}} L\right) - 1\right)},\\
    \hat{C}_D(s) &= e^{-\sqrt{\frac{s+r}{D_0}} L}\frac{1+ \hat{J}_R^{[2]}(L^{-},s) - \hat{J}_R^{[1]}(0^{+},s)}{2\sqrt{D_0(r+s)}\left(\exp\left(\sqrt{\frac{s+r}{D_0}} L\right) - 1\right)}.
\end{align}
\end{subequations}
Putting these together and defining
\begin{equation}\label{eq:ExpressionNormalisationConstant_Appendix}
     \hat{A}_D(s) = \frac{1+\hat{J}_R^{[2]}(L^{-},s) - \hat{J}_R^{[1]}(0^{+},s)}{2\sqrt{D_0 (r+s)}},
\end{equation}
we finally obtain for the Laplace-transformed diffusion-phase density, 
\begin{equation}\label{eq:DiffusionLaplaceSolution}
     \hat{P}_D(x,s) = \hat{A}_D(s) \frac{\cosh\left( \sqrt{\frac{s+r}{D_0}} \left(x - \frac{L}{2} \right) \right)}{\sinh\left( \sqrt{\frac{s+r}{D_0}}\frac{L}{2} \right)}.
\end{equation}

We now proceed to obtain the time-dependent solution for the resetting-phase probability density $P_R(x,t)$, described by Eqs.~(\ref{subeq:FP_resetting_appendix1})-(\ref{subeq:FP_resetting_appendix2}). Taking the Laplace transform of Eqs.~(\ref{subeq:FP_resetting_appendix1})-(\ref{subeq:FP_resetting_appendix2}), we obtain
\begin{subequations}
    \label{eq:LaplacePDEResetting}
    \begin{align}
        D_0 \frac{\partial^2 \hat{P}_R^{[1]}(x,s)}{\partial x^2} + V'^{[1]}\frac{\partial \hat{P}_R^{[1]}(x,s)}{\partial x} -s\hat{P}_R^{[1]}(x,s) + r\hat{P}_D(x,s) + \hat{J}_R^{[1]}(0^{+},s)\delta(x) &= 0,\label{eq:LaplacePDEResetting1}\\
        D_0 \frac{\partial^2 \hat{P}_R^{[2]}(x,s)}{\partial x^2} + V'^{[2]}\frac{\partial \hat{P}_R^{[2]}(x,s)}{\partial x} -s\hat{P}_R^{[2]}(x,s) + r\hat{P}_D(x,s) - \hat{J}_R^{[2]}(L^{-},s) \delta(x) &= 0. \label{eq:LaplacePDEResetting2}
    \end{align}
\end{subequations}
As for the diffusion-phase density, we consider Eq.~(\ref{eq:LaplacePDEResetting}) in the region $x \in (0,L)$ such that, after inserting the solution for $\hat{P}_D(x,s)$, Eq.~(\ref{eq:DiffusionLaplaceSolution}), we obtain the following partial differential equations for each component of the Laplace-transformed resetting-phase density,
\begin{subequations}
    \label{eq:ResettingPhasePDEs}
    \begin{align}
         D_0\frac{\partial^2 \hat{P}_R^{[1]}(x,s)}{\partial x^2} + \frac{h}{a}\frac{\partial \hat{P}_R^{[1]}(x,s)}{\partial x} - s \hat{P}_R^{[1]}(x,s) +r\hat{A}_D(s)\frac{\cosh\left( \sqrt{\frac{s+r}{D_0}} \left(x - \frac{L}{2} \right) \right)}{\sinh\left( \sqrt{\frac{s+r}{D_0}}\frac{L}{2} \right)} &= 0 ,\label{pdeR1} \\
         D_0\frac{\partial^2 \hat{P}_R^{[2]}(x,s)}{\partial x^2} - \frac{h}{L - a}\frac{\partial \hat{P}_R^{[2]}(x,s)}{\partial x} - s \hat{P}_R^{[2]}(x,s) +r\hat{A}_D(s)\frac{\cosh\left( \sqrt{\frac{s+r}{D_0}} \left(x - \frac{L}{2} \right) \right)}{\sinh\left( \sqrt{\frac{s+r}{D_0}}\frac{L}{2} \right)} &= 0 .\label{pdeR2}
    \end{align}
\end{subequations}
The solutions for each component of the resetting-phase density are then given by
\begin{subequations}\label{eq:ResettingPhaseSolution_Appendix}
    \begin{align}
        \hat{P}_R^{[1]}(x,s) &= \hat{g}^{[1]}(x,s) + \hat{A}_R^{[1]}(s) \hat{h}_{a_{1}}(x,s) + \hat{B}_R^{[1]}(s) \hat{h}_{b_{1}}(x,s), \\
        \hat{P}_R^{[2]}(x,s) &= \hat{g}^{[2]}(x,s) + \hat{A}_R^{[2]}(s) \hat{h}_{a_{2}}(x,s) + \hat{B}_R^{[2]}(s) \hat{h}_{b_{2}}(x,s).
    \end{align}
\end{subequations}
where the $\hat{g}^{[i]}(x,s)$ are the particular solutions of Eq.~(\ref{eq:ResettingPhasePDEs}), $(\hat{h}_{a_i}, \hat{h}_{b_i})$ are the homogeneous solutions, and $(\hat{A}^{[i]}_R, \hat{B}^{[i]}_R)$ are constants to be fixed. The homogeneous solutions are given by
\begin{subequations}
    \begin{align}
        \hat{h}_{a_{1}}(x,s) &= \exp\left( -\frac{x}{2D_0a} \left( h + \sqrt{h^2 + 4 D_0 a^2 s} \right) \right),\\
        \hat{h}_{b_{1}}(x,s) &= \exp\left( -\frac{x}{2D_0a} \left( h - \sqrt{h^2 + 4 D_0 a^2 s} \right) \right), \\
        \hat{h}_{a_{2}}(x,s) &= \exp\left( \frac{x}{2D_0(L-a)} \left( h + \sqrt{h^2 + 4 D_0 a^2 s} \right) \right), \\
        \hat{h}_{b_{2}}(x,s) &= \exp\left( \frac{x}{2D_0(L-a)} \left(h - \sqrt{h^2 + 4 D_0 a^2 s} \right) \right),
    \end{align}
\end{subequations}
while the particular solutions (for the case when both $h/a \neq \sqrt{Dr}$ and $h/(L-a) \neq \sqrt{Dr}$) are given by
\begin{subequations}
    \begin{align}
        \hat{g}^{[1]}(x,s) &= \hat{A}_D(s) \frac{D_0 r a^2}{(r+s)h^2 - D_0 r a^2} \frac{r \cosh\left( \sqrt{\frac{r+s}{D_0}}\left(x-\frac{L}{2}\right) \right) - \sqrt{\frac{r+s}{D_0}} \frac{h}{a} \sinh\left( \sqrt{\frac{r+s}{D_0}}\left(x-\frac{L}{2}\right)\right)}{\sinh\left(\sqrt{\frac{r+s}{D_0}} \frac{L}{2}\right)},\\
        \hat{g}^{[2]}(x,s) &=  \hat{A}_D(s) \frac{D_0 r (L-a)^2}{(r+s)h^2 - D_0 r (L-a)^2} \frac{r \cosh\left( \sqrt{\frac{r+s}{D_0}}\left(x-\frac{L}{2}\right) \right) + \sqrt{\frac{r+s}{D_0}} \frac{h}{L -a} \sinh\left( \sqrt{\frac{r+s}{D_0}}\left(x-\frac{L}{2}\right)\right)}{\sinh\left(\sqrt{\frac{r+s}{D_0}} \frac{L}{2}\right)}.
    \end{align}
\end{subequations}

Next, we fix the constants $\hat{A}^{[1]}_R (s), \hat{A}^{[2]}_R (s), \hat{B}^{[1]}_R (s), \hat{B}^{[2]}_R (s)$ in Eq.~(\ref{eq:ResettingPhaseSolution_Appendix}). In addition to the periodic boundary condition, Eq.~(\ref{eq:PBCappendixResetting}), we also have the following continuity conditions at the apex $x=a$,
\begin{subequations}\label{eq:Continuity_dimensionless_Resetting_appendix}
    \begin{align}
    P_R^{[1]}(a,t) &= P_R^{[2]}(a,t), \label{subeq:Continuity_Resetting_dimensionless_appendix}\\
    J_R^{[1]}(a,t) &= J_R^{[2]}(a,t), \label{subeq:CurrentContinuity_Resetting_dimensionless_appendix}
    \end{align}
\end{subequations}
where the latter condition (\ref{subeq:CurrentContinuity_Resetting_dimensionless_appendix}) can be expressed explicitly as the jump condition
\begin{equation}
\label{jumpcondition_R}
    D_0\left( \left. \frac{\partial P_R^{[2]}(x,t)}{\partial x}\right|_{x \to a^+} - \left. \frac{\partial P_R^{[1]}(x,t)}{\partial x}\right|_{x \to a^-}\right)
    = \left( V'^{[1]}(x\to a^{-}) - V'^{[2]}(x \to a^{+})\right)P_R^{[1]}(a,t).
\end{equation}

The boundary conditions for the resetting-phase density $\hat{P}_R^{[i]}(x,s)$ can then be summarised by the matrix equation $\mathsf{M}(s) \cdot \left(A_R^{[1]}(s), B_R^{[1]}(s), A_R^{[2]}(s), B_R^{[2]}(s)\right) = \textbf{u}(s)$, explicitly
\begin{equation}\label{eq:BCsMatrixEq_Appendix}
\begin{aligned}
     & \begin{bmatrix}
    \hat{h}_{a_1}(0,s) & \hat{h}_{b_1}(0,s) & 0 & 0\\
    0 & 0 & \hat{h}_{a_2}(L,s) & \hat{h}_{b_2}(L,s) \\
    \hat{h}_{a_1}(a,s)& \hat{h}_{b_1}(a,s) & \hat{h}_{a_2}(a,s) & \hat{h}_{b_2}(a,s) \\
    \frac{h/a + h/(L-a)}{D_0}\hat{h}_{a_1}(a,s) + \partial_x \hat{h}_{a_1}(a,s) & \frac{h/a + h/(L-a)}{D_0}\hat{h}_{b_1}(a,s) + \partial_x \hat{h}_{b_1}(a,s) & -\partial_x \hat{h}_{a_2}(a,s) & -\partial_x \hat{h}_{b_2}(a,s) \\
    \end{bmatrix}
    \cdot 
    \begin{bmatrix}
    \hat{A}_R^{[1]}(s) \\
    \hat{B}_R^{[1]}(s) \\
    \hat{A}_R^{[2]}(s) \\
    \hat{B}_R^{[2]}(s)
    \end{bmatrix}
    = \\
    &
    \begin{bmatrix}
    -\hat{g}^{[1]}(0,s) \\
    -\hat{g}^{[2]}(L,s) \\
    -\left( \hat{g}^{[1]}(a,s) - \hat{g}^{[2]}(a,s)\right) \\
    -\left( \partial_x \hat{g}^{[1]}(a,s) - \partial_x \hat{g}^{[2]}(a,s) + \frac{h/a + h/(L-a)}{D_0} \hat{g}^{[1]}(a,s)\right)
    \end{bmatrix},
\end{aligned}
\end{equation}
where the expressions arising from the first two rows of Eq.~(\ref{eq:BCsMatrixEq_Appendix}) enforce the periodic boundary condition (\ref{eq:PBCappendixResetting}), the third row enforces Eq.~(\ref{subeq:Continuity_Resetting_dimensionless_appendix}), and the final row enforces Eq.~(\ref{subeq:CurrentContinuity_Resetting_dimensionless_appendix}). We can obtain explicit expressions for the constants $\hat{A}^{[1]}_R(s), \hat{A}^{[2]}_R(s), \hat{B}^{[1]}_R(s), \hat{B}^{[2]}_R(s)$ in terms of the system parameters by multiplying the matrix inverse with the vector on the right-hand side of Eq.~(\ref{eq:BCsMatrixEq_Appendix}), i.e.\ $\left(\hat{A}_R^{[1]}(s), \hat{B}_R^{[1]}(s), \hat{A}_R^{[2]}(s), \hat{B}_R^{[2]}(s)\right) = \mathsf{M}^{-1}(s) \cdot \textbf{u}(s)$. It then remains to find an explicit expression for $\hat{J}_R^{[2]}(L^{-},s) - \hat{J}_R^{[1]}(0^{+},s)$ in terms of the system parameters. This can be achieved using the jump condition of the resetting-phase density at the origin, obtained by integrating Eq.~(\ref{eq:LaplacePDEResetting}) over the region $[-\epsilon, \epsilon]$, recalling $-\epsilon \sim L - \epsilon$, and then taking $\epsilon \to 0^+$, i.e.\
\begin{equation}
\label{eq:JumpConditionResettingOrigin_Appendix}
    D_0\left( \left. \frac{\partial \hat{P}_R^{[1]}(x,s)}{\partial x}\right|_{x \to 0^{+}} - \left. \frac{\partial \hat{P}_R^{[2]}(x,s)}{\partial x}\right|_{x \to L^{-}}\right) = \hat{J}_R^{[2]}(L^{-},s) - \hat{J}_R^{[1]}(0^{+},s) ,
\end{equation}
which concludes the derivation of the solutions for the densities in Laplace space.

In principle, exact expressions for the time-dependent probability densities in real space, i.e.\ $P_{D,R}(x,t)$, can be obtained by taking the inverse Laplace transform of Eqs.~(\ref{eq:DiffusionLaplaceSolution}) and (\ref{eq:ResettingPhaseSolution_Appendix}), though the resulting expressions are too complicated to write down here explicitly. However, we can recover the steady-state limits of the probability densities $P_{D,R}(x)$ derived in the main text, Eqs.~(\ref{eq:ProbabilityDensity_Diffusion_Solution}) and (\ref{eq:ProbabilityDensity_Resetting_Solution}), through the final value theorem, i.e.\
\begin{equation}
\label{eq:FinalValueTheorem}
    \lim_{t \to \infty} P_{D,R}(x,t) = \lim_{s \to 0}s P_{D,R}(x,s).
\end{equation}
It turns out that $\lim_{s \to0} s \hat{A}_D(s) = p_D/2$, where $\hat{A}_D(s)$ is given by Eq.~(\ref{eq:ExpressionNormalisationConstant_Appendix}) and $p_D$ is the normalisation of the diffusion-phase probability density, defined by Eq.~(\ref{eq:ProbabilityDiffusionPhase}) in the main text. For the case when both $h/a \neq \sqrt{Dr}$ and $h/(L-a) \neq \sqrt{Dr}$, we recover the steady-state solutions of the main text, Eqs.~(\ref{eq:ProbabilityDensity_Diffusion_Solution}) and (\ref{eq:ProbabilityDensity_Resetting_Solution}), namely (in dimensionless form),
\begin{subequations}
\begin{align}
    \tilde{P}_D(\tilde{x}) &= \frac{p_D}{2}\frac{\cosh(\tilde{x}-\frac{\ell}{2})}{\sinh(\frac{\ell}{2})}, \label{eq:ProbabilityDensity_Diffusion_Solution_General_Appendix}\\
    \tilde{P}_R(\tilde{x}) &= 
    \begin{cases}
      \frac{p_D}{2} \frac{ \delta \ell}{e^{\ell} - 1}\left( \frac{e^{\ell - \tilde{x}}}{\lambda^{[1]}_{-}} - \frac{e^{\tilde{x}}}{\lambda^{[1]}_{+}}\right) + A_R^{[1]} e^{-\frac{\kappa \tilde{x}}{ \delta \ell}} + B_R^{[1]}, & 0 \leq \tilde{x} < \delta\ell, \\
      \frac{p_D}{2} \frac{ (1 - \delta) \ell}{e^{\ell} - 1}\left( \frac{e^{\tilde{x}}}{\lambda^{[2]}_{-}} - \frac{e^{\ell - \tilde{x}}}{\lambda^{[2]}_{+}}\right) + A_R^{[2]} e^{\frac{\kappa \tilde{x} }{ (1 - \delta) \ell}} + B_R^{[2]},  & \delta\ell \leq \tilde{x} < \ell,
    \end{cases}\label{eq:ProbabilityDensity_Resetting_Solution_General_Appendix}
\end{align}
\end{subequations}
    \begin{subequations}
    \end{subequations}

For completeness, the edge-case solutions for the resetting-phase probability density are given by 
\begin{itemize}
    \item \underline{Case (i): $ h/a = \sqrt{Dr} \neq h/(L-a)$, $\left(\lambda_{-}^{[1]} = 0 \neq \lambda_{-}^{[2]} \right)$}
    \begin{subequations}
    \begin{align}   
    \tilde{P}_R(\tilde{x}) &= 
    \begin{cases*}
      \frac{p_D}{2} \frac{ 2e^{\ell - \tilde{x}} (1+\tilde{x}) -e^{\tilde{x}}}{2(e^{\ell} - 1)} + A_R^{[1]} e^{-\tilde{x}} + B_R^{[1]}, & $0 \leq \tilde{x} < \delta\ell$, \\
      \frac{p_D}{2} \frac{\ell (1 - \delta)}{e^{\ell} - 1}\left( \frac{e^{\tilde{x}}}{\lambda^{[2]}_{-}} - \frac{e^{\ell - \tilde{x}}}{\lambda^{[2]}_{+}}\right) + A_R^{[2]} e^{\frac{\tilde{x} \kappa}{\ell (1 - \delta)}} + B_R^{[2]},  & $\delta\ell \leq \tilde{x} < \ell$,
    \end{cases*}
    \end{align} 
    \item \underline{Case (ii): $h/(L-a) = \sqrt{Dr} \neq h/a$, $\left(\lambda_{-}^{[2]} = 0 \neq \lambda_{-}^{[1]} \right)$}
    \begin{align}   
    \tilde{P}_R(\tilde{x}) &= 
    \begin{cases*}
      \frac{p_D}{2} \frac{\ell \delta}{e^{\ell} - 1}\left( \frac{e^{\ell - \tilde{x}}}{\lambda^{[1]}_{-}} - \frac{e^{\tilde{x}}}{\lambda^{[1]}_{+}}\right) + A_R^{[1]} e^{-\frac{\tilde{x} \kappa}{\ell \delta}} + B_R^{[1]}, & $0 \leq \tilde{x} < \delta\ell$, \\
      \frac{p_D}{2} \frac{ -2e^{\tilde{x}} (1-\tilde{x}) + e^{\ell - \tilde{x}}}{2(e^{\ell} - 1)} + A_R^{[2]} e^{\tilde{x}} + B_R^{[2]},  & $\delta\ell \leq \tilde{x} < \ell$,
    \end{cases*}
    \end{align}
    \item \underline{Case (iii): $h/a = h/(L-a) = \sqrt{Dr}$, $\left(\lambda_{-}^{[1]} = \lambda_{-}^{[2]} = 0 \right)$}
    \begin{align}   
    \tilde{P}_R(\tilde{x}) &= 
    \begin{cases*}
      \frac{p_D}{2} \frac{ 2e^{\ell - \tilde{x}} (1+\tilde{x}) -e^{\tilde{x}}}{2(e^{\ell} - 1)} + A_R^{[1]} e^{-\tilde{x}} + B_R^{[1]}, & $0 \leq \tilde{x} < \delta\ell$, \\
     \frac{p_D}{2} \frac{ -2e^{\tilde{x}} (1-\tilde{x}) + e^{\ell - \tilde{x}}}{2(e^{\ell} - 1)} + A_R^{[2]} e^{\tilde{x}} + B_R^{[2]}, & $\delta\ell \leq \tilde{x} < \ell$,
    \end{cases*}
    \end{align}
\end{subequations}
\end{itemize}
where the constants $A^{[1]}_R, A^{[2]}_R, B^{[1]}_R, B^{[2]}_R$ differ for each case, and can be calculated using the final value theorem, e.g.\ $A^{[1]}_R = \lim_{s\to 0} s\hat{A}^{[1]}_R(s)$.

%%%%%%%%%%%%%%%%%%%%%%%%%%%%%%%%%%%%%%%%%%%%%%%%%%%%%%%%%

\section{Solution for a ``stop-and-start" model with deterministic returns in a static ratchet} \label{App:Sec:DeterministicRatchetModel}
In this appendix, we discuss how to obtain the steady-state solution for a ``stop-and-start" resetting model. In this model, while diffusing in a ratchet potential, a particle stochastically switches off its diffusion. This commences the ``resetting phase", which ends only upon the particle reaching the origin, where it then switches its diffusion back on again, repeating the cycle. Since the particle does not diffuse during the resetting phase, the returns to the origin are deterministic. The dynamics are captured by the following Langevin equations,
\begin{subequations}
\begin{align}
    \dot{x}_D(t) &= -V'(x_D(t)) + \sqrt{2D_0} \xi(t), \\
    \dot{x}_R(t) & = -V'(x_R(t)),
\end{align}
\end{subequations}
where the ratchet potential $V(x)$ is the same as the one defined in Eq.~(\ref{eq:ratchetpotential}). This model is similar to that considered in Ref.~\cite{ghosh2023}, for which the authors obtain numerical results. However, in that work, the ratchet potential is continuously differentiable, whereas we will continue to consider a piecewise-linear ratchet potential, Eq.~(\ref{eq:ratchetpotential}).

\begin{figure}
    \centering
    \includegraphics[width=0.5\linewidth]{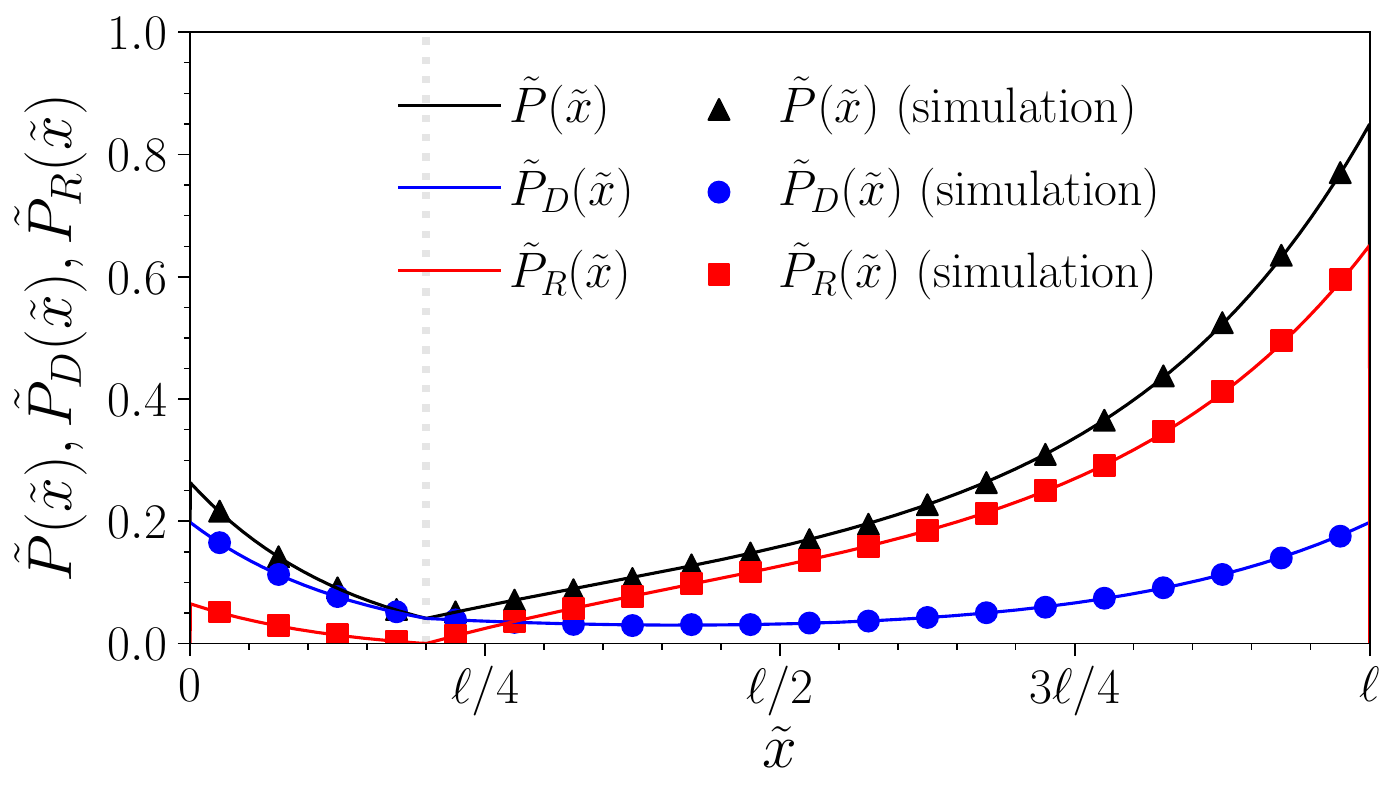}
    \caption{Steady-state probability densities $\tilde{P}_{D,R}(\tilde{x})$ for the ``stop-and-start" model as a function of particle position $\tilde{x}$ for parameters $\ell=4$, $\kappa=1$, and $\delta=0.2$. The vertical dotted line indicates the position of the ratchet apex $\tilde{x} = \delta\ell$. Simulation data are plotted as symbols and are in excellent agreement with the theoretical results obtained by the procedure described in Appendix~\ref{App:Sec:DeterministicRatchetModel}.}
    \label{fig:ProbabilityDensitiesStopStartVsPosition}
\end{figure}

One way to solve this model is to first consider that the diffusion and resetting phases each have an associated finite diffusion constant, say $D_D$ and $D_R$, respectively. After solving this more general model, one can take the limit $D_R \to 0$ to recover the stop-and-start model. It turns out that solving the model in this way avoids an issue with solving the $D_R = 0$ case directly, namely that the usual resetting-phase boundary conditions $P_R(0) = P_R(L) = 0$ no longer hold for $D_R = 0$ as the resetting-phase density has a finite value at the origin in this case \cite{pal2019, pal2019time}. Solving the more general model for finite $D_R$ allows us to keep the usual resetting-phase boundary conditions. However, one needs to take some care as the limits $x \to 0$ and $D_R \to 0$ do not commute \cite{gupta2021}. Hence, taking the limit $D_R \to 0$ in the more general model recovers the $D_R = 0$ result everywhere except at $x = 0$, where the solution abruptly drops to $0$ due to the boundary conditions.

The steady-state Fokker-Planck equations for the model with finite $D_R$ can be cast in the following dimensionless form,
\begin{subequations}
    \label{eq:FPequation_StopStartModel_nondimensional}
    \begin{align}
        0 &= \frac{\partial^2 \tilde{P}^{[1]}_D(\tilde{x})}{\partial \tilde{x}^2} + \tilde{V}'^{[1]}\frac{\partial \tilde{P}^{[1]}_D(\tilde{x}) }{\partial \tilde{x}}- \tilde{P}^{[1]}_D(\tilde{x}) - \tilde{J}_R^{[1]}(0^{+})\delta(\tilde{x}) ,\label{eq:FP_Diffusion_StopStart_NonDim1} \\
        0 &= \frac{\partial^2 \tilde{P}^{[2]}_D(\tilde{x})}{\partial \tilde{x}^2} + \tilde{V}'^{[2]}\frac{\partial \tilde{P}^{[2]}_D(\tilde{x}) }{\partial \tilde{x}}- \tilde{P}^{[2]}_D(\tilde{x}) + \tilde{J}_R^{[2]}(\ell^{-})\delta(\tilde{x}) ,\label{eq:FP_Diffusion_StopStart_NonDim2} \\
         0 &= \alpha \frac{\partial^2 \tilde{P}^{[1]}_R(\tilde{x})}{\partial \tilde{x}^2} +  \tilde{V}'^{[1]}\frac{\partial \tilde{P}^{[1]}_R(\tilde{x})}{\partial \tilde{x}} +  \tilde{P}_D^{[1]}(\tilde{x}) + \tilde{J}_R^{[1]}(0^{+})\delta(\tilde{x}) , \label{eq:FP_Resetting_StopStart_NonDim1} \\
         0 &= \alpha \frac{\partial^2 \tilde{P}^{[2]}_R(\tilde{x})}{\partial \tilde{x}^2} +  \tilde{V}'^{[2]}\frac{\partial \tilde{P}^{[2]}_R(\tilde{x})}{\partial \tilde{x}} +  \tilde{P}_D^{[2]}(\tilde{x}) -  \tilde{J}_R^{[2]}(\ell^{-})\delta(\tilde{x}) , \label{eq:FP_Resetting_StopStart_NonDim2}
    \end{align}
\end{subequations}
where we have introduced $\alpha = D_R/D_0$ as a new dimensionless parameter and all other symbols retain the same definitions from the main text. One can verify that the Boltzmann solution, i.e.\ $\tilde{P}(\tilde{x}) = \tilde{P}_R(\tilde{x}) + \tilde{P}_D(\tilde{x}) \propto \exp(-\tilde{V}(\tilde{x}))$ is recovered for $\alpha = 1$. The stop-and-start model is recovered in the limit $\alpha \to 0$.

We now describe how to obtain the solution to Eq.~(\ref{eq:FPequation_StopStartModel_nondimensional}). Since the procedure is mostly analogous to Appendix~\ref{App:Sec:DensitySolutions}, we will omit most of the details. As in Appendix~\ref{App:Sec:DensitySolutions}, we first solve for the diffusion-phase density, and then fix the constants of integration using the periodic boundary conditions $\tilde{P}_D^{[1]}(0) = \tilde{P}_D^{[2]}(\ell)$, density continuity condition $\tilde{P}_D^{[1]}(\delta\ell) = \tilde{P}_D^{[2]}(\delta\ell)$, current continuity condition $\tilde{J}_D^{[1]}(\delta\ell) = \tilde{J}_D^{[2]}(\delta\ell)$, and jump condition at $\tilde{x} = 0$, which can be found by integrating Eqs.~(\ref{eq:FP_Diffusion_StopStart_NonDim1})-(\ref{eq:FP_Diffusion_StopStart_NonDim2}) over the region $\tilde{x} \in [-\epsilon, \epsilon]$, recalling $-\epsilon \sim \ell-\epsilon$, and then taking $\epsilon \to 0^+$, resulting in
\begin{equation}
    \label{eq:modified_jump_condition}
    \left. \frac{\partial \tilde{P}_D^{[1]}(\tilde{x}) }{\partial \tilde{x}} \right|_{\tilde{x} \to 0^+} - \left .\frac{ \partial \tilde{P}_D^{[2]}(\tilde{x})}{\partial \tilde{x}}\right|_{\tilde{x} \to \ell^-} + \frac{\kappa}{ \ell \delta (1 - \delta) }\tilde{P}^{[1]}_D(0) = -\left(\tilde{J}_R^{[2]}(\ell^{-}) - \tilde{J}_R^{[1]}(0^{+})\right),
\end{equation}
where $\tilde{J}_R^{[2]}(\ell^{-}) - \tilde{J}_R^{[1]}(0^{+})$ can be fixed using the normalisation condition $p_D + p_R = 1$. The solution for $\tilde{P}_D(\tilde{x})$ can then be inserted into Eqs.~(\ref{eq:FP_Resetting_StopStart_NonDim1})-(\ref{eq:FP_Resetting_StopStart_NonDim2}) to solve for $\tilde{P}_R(\tilde{x})$, the procedure of which is analogous to the above. The resulting expressions for the densities of both phases are too complicated to write down here.
However, we have plotted the densities in figure~\ref{fig:ProbabilityDensitiesStopStartVsPosition}, where the analytic expressions obtained by the procedure above are shown to be in good agreement with simulations.

%%%%%%%%%%%%%%%%%%%%%%%%%%%%%%%%%%%%%%%%%%%%%%%%%%%%%%%%%%%%%%%%%%%

\section{Relation between normalisation conditions, Eqs.~(\ref{eq:NormalisationCondition}) and (\ref{eq:NormalisationCondition_MFPT})}\label{app:sec:RelatingNormalisationConditions}

In this appendix, we show that Eq.~(\ref{eq:NormalisationCondition_MFPT}) can be derived from Eq.~(\ref{eq:NormalisationCondition}) without the need for an ad-hoc justification based solely on physical arguments.

We begin by multiplying both sides of Eq.~(\ref{eq:NormalisationCondition}), $p_D + p_R = 1$, by $\tau$ and then rearranging to obtain
\begin{equation}
\label{eq:secondstep}
    p_D = \frac{\tau}{\tau + \frac{\tau}{p_D} p_R}.
\end{equation}
Next, we derive an expression for $p_R$ in terms of the mean first-passage time $T_R(x)$. First, we multiply the dimensionless Fokker-Planck equation (\ref{subeq:FP_resetting_dimensionless1})-(\ref{subeq:FP_resetting_dimensionless2}) by the dimensionless mean first-passage time $\tilde{T}_R(x)$, and then integrate over the entire region. This results in
\begin{subequations}
\begin{align}
\label{eq:trick}
    0 &= \int_{0}^{\ell} d\tilde{x}~\tilde{T}_R(x) \left[ \frac{\partial^2 \tilde{P}_R(\tilde{x})}{\partial \tilde{x}^2}  + \frac{\partial}{\partial \tilde{x}}\left(\tilde{V}'(\tilde{x}) \tilde{P}_R(\tilde{x})\right) + \tilde{P}_D(\tilde{x}) - \left(\tilde{J}_R(\ell^{-}) - \tilde{J}_R(0^{+})\right)\delta(\tilde{x}) \right]  \\
     &= \int_{0}^{\ell} d\tilde{x}~\left[ \frac{\partial^2 \tilde{T}_R(\tilde{x})}{\partial \tilde{x}^2}  - \tilde{V}'(\tilde{x})\frac{\partial \tilde{T}_R(\tilde{x})}{\partial \tilde{x}} \right] \tilde{P}_R(\tilde{x}) + \int_0 ^{\ell} d\tilde{x}~\tilde{P}_D(\tilde{x}) \tilde{T}_{R}(\tilde{x}) \label{eq:trick2}
    \end{align}
\end{subequations}
where, in the second equality, we integrated by parts twice with the resulting boundary terms vanishing due to the absorbing boundary conditions satisfied by $\tilde{P}_R(\tilde{x})$ and $\tilde{T}_R(\tilde{x})$. The term proportional to $\delta(\tilde{x})$ also vanishes because $\tilde{T}_R(0) = 0$. The first integral on the right-hand side of Eq.~(\ref{eq:trick2}) can be simplified using the definition of the mean first-passage time given by Eq.~(\ref{eq:MFPT_ResettingPhase}), i.e.\
\begin{equation}
    \frac{\partial^2 \tilde{T}_R(\tilde{x})}{\partial \tilde{x}^2}  - \tilde{V}'(\tilde{x})\frac{\partial \tilde{T}_R(\tilde{x})}{\partial \tilde{x}} = -1.
    \end{equation}
Hence, the first integral on the right-hand side of Eq.~(\ref{eq:trick2}) simplifies drastically to $-\int_0^\ell d\tilde{x}~\tilde{P}_R(\tilde{x}) = -p_R$, whence we obtain
\begin{equation}\label{eq:pR_Expression_appendix}
    p_R = \int_0 ^{\ell} d\tilde{x}~\tilde{P}_D(\tilde{x}) \tilde{T}_{R}(\tilde{x}).
\end{equation}
Substituting this expression for $p_R$, Eq.~(\ref{eq:pR_Expression_appendix}), into Eq.~(\ref{eq:secondstep}) (which is just a rearrangement of Eq.~(\ref{eq:NormalisationCondition})) recovers Eq.~(\ref{eq:NormalisationCondition_MFPT}) in the main text, thus concluding the derivation.

%%%%%%%%%%%%%%%%%%%%%%%%%%%%%%%%%%%%%%%%%%%%%%%%%%%%%%%%%%%%%%%%

\section{Mean first-passage time in the resetting phase} \label{App:Sec:ResettingMFPT}
In this appendix, we derive the (unconditional) mean first-passage time (MFPT) $T_{R}(x_0)$ of the particle in the resetting phase, i.e.\ the average time taken for a particle initialised at position $x_0 \in [0,L]$ in the resetting phase to exit the resetting phase by reaching either $x=0$ or $x=L$. To be consistent with the main text, we define our units of length and time as $\sigma = \sqrt{D_0/r}$ and $\tau = 1/r$, respectively, despite the MFPT being independent of $r$. While the solution we derive below depends on the dimensionless parameters $\kappa$, $\delta$ and $\ell$, the latter of which depends on $r$, the dependence on $r$ vanishes after converting back to dimensionful parameters, as should be expected.

\begin{figure}
    \centering
    \includegraphics[width=0.5\linewidth]{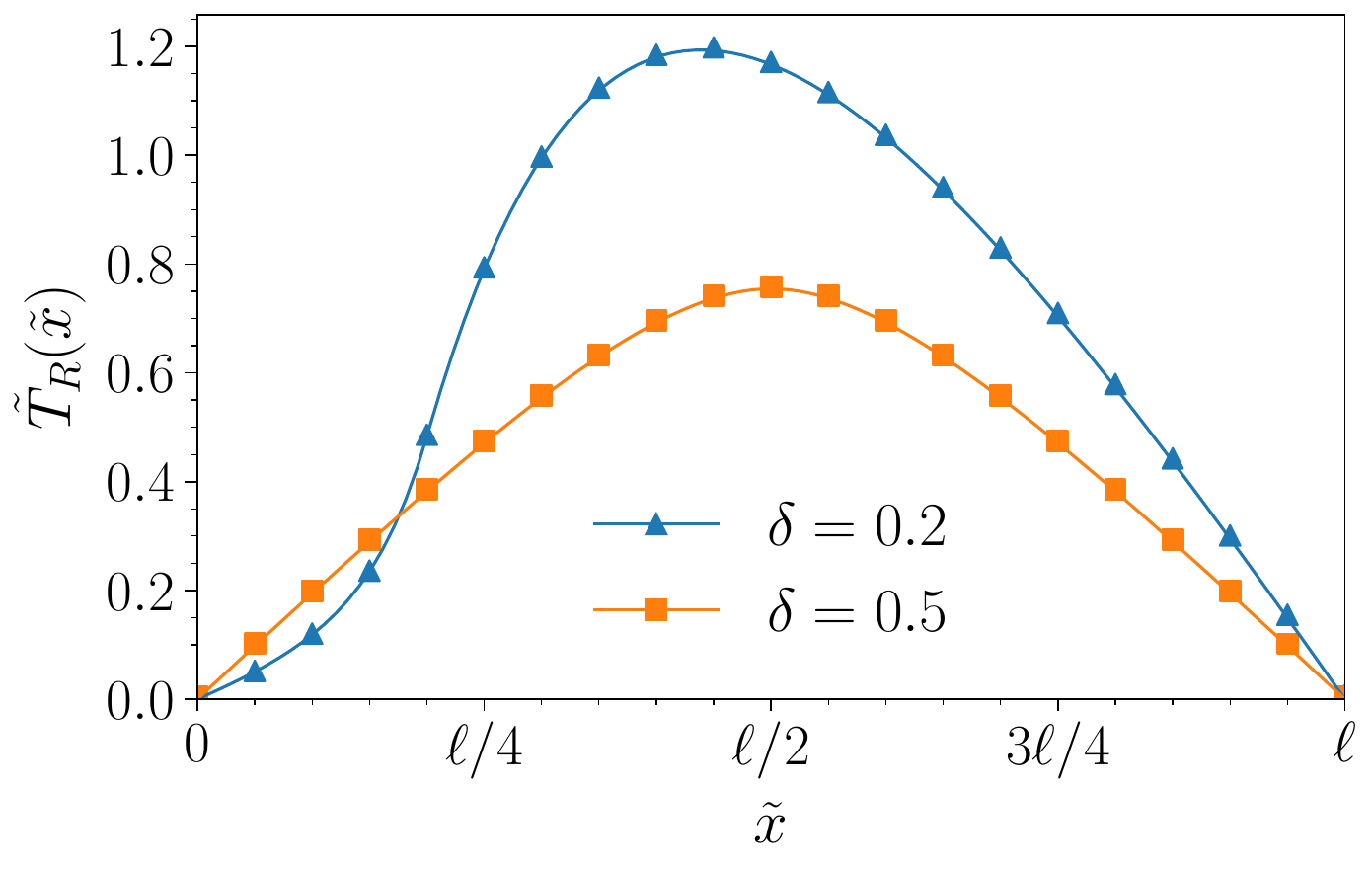}
    \caption{Mean first-passage time $\tilde{T}_R(\tilde{x})$ to exit the resetting phase, Eq.~(\ref{eq:MFPT_FinalExpression}), as a function of $\tilde{x}$ for $\ell=4$, $\kappa=4$, and $\delta=0.2$ (blue) or $\delta=0.5$ (orange). Simulation results are given by markers and are obtained by averaging over $10^6$ realizations for each initial position $\tilde{x}/\ell = 0, 0.05, \dots, 1$.}
    \label{fig:MFPT_vs_x}
\end{figure}

In dimensionless form, the MFPT obeys the following partial differential equation \cite{van2007},
\begin{equation}
    \label{eq:MFPT_ResettingPhase}
         -1 = \frac{\partial^2 \tilde{T}_R(\tilde{x})}{\partial \tilde{x}^2}  - \tilde{V}'(\tilde{x})\frac{\partial \tilde{T}_R(\tilde{x})}{\partial \tilde{x}},
\end{equation}
where the advective term has the opposite sign compared to that of the probability densities (see, for example, Eqs.~(\ref{subeq:FP_resetting_dimensionless1})-(\ref{subeq:FP_resetting_dimensionless2})), due to the MFPT obeying the backwards Kolmogorov equation. By definition, we have the following absorbing boundary conditions
\begin{equation}
    \tilde{T}_R^{[1]}(0) = \tilde{T}_R^{[2]}(\ell) = 0,
\end{equation}
which impose that a particle initialised at either $\tilde{x} = 0$ or $\tilde{x} = \ell$ is absorbed instantly and thus takes zero time to exit the resetting phase. The MFPT is also subject to the following continuity conditions at the apex $\tilde{x} = \delta\ell$ \cite{roberts2023},
\begin{subequations}
\begin{align}
    \tilde{T}_R^{[1]}(\delta \ell) &= \tilde{T}_R^{[2]}(\delta \ell),\\
    \left. \frac{\partial \tilde{T}_R^{[1]}(\tilde{x})}{\partial \tilde{x}} \right|_{\tilde{x} \to \delta \ell^-} &= \left. \frac{\partial \tilde{T}_R^{[2]}(\tilde{x})}{\partial \tilde{x}} \right|_{\tilde{x} \to \delta \ell^+}.
\end{align}
\end{subequations}
In a similar manner to that of the probability densities in Appendix~\ref{App:Sec:DensitySolutions}, we can solve Eq.~(\ref{eq:MFPT_ResettingPhase}), whence we obtain
\begin{small}
\begin{equation}\label{eq:MFPT_FinalExpression}
    \tilde{T}_R(\tilde{x}) = 
    \begin{cases*}
      \frac{\ell \delta}{\kappa^2(e^{\kappa} - 1)}\left( \ell (1 - \delta)e^{-\kappa}(e^{\frac{\kappa}{\ell \delta} \tilde{x}} - 1) + \kappa \tilde{x}(e^{\kappa} - 1) - \ell (e^{\frac{\kappa}{\ell \delta} \tilde{x}} - 1)(1- \kappa +\delta (2\kappa-1)) \right), & $0 \leq \tilde{x} < \delta \ell$, \\
      \frac{\ell (1 - \delta)}{\kappa^2(e^{\kappa} - 1)}\left( \ell \delta \left[ e^{-\kappa \frac{\tilde{x}/\ell - \delta}{1 - \delta } } - e^{-\kappa}\right] + e^{\kappa}\kappa (\ell - \tilde{x}) - \ell e^{\kappa - \frac{\kappa}{1 - \delta}(\tilde{x}/\ell - \delta)}(\delta + \kappa -2\delta \kappa) + (\tilde{x} \kappa + \ell \delta(1 - 2\kappa)) \right), & $\delta \ell \leq x < \ell$.
    \end{cases*}
\end{equation}
\end{small}
The analytical expression for the MFPT, Eq.~(\ref{eq:MFPT_FinalExpression}), is plotted in figure~\ref{fig:MFPT_vs_x} and shows good agreement with simulation results.

%%%%%%%%%%%%%%%%%%%%%%%%%%%%%%%%%%%%%%%%%%%%%%%%%%%%%%%%%
%
\bibliography{references}
\end{document}